\documentclass[a4paper,10pt]{article}
\pdfoutput=1 

\usepackage{jheppub} 

\usepackage[T1]{fontenc} 

\usepackage{setspace}

\usepackage{physics}
\usepackage{dsfont}

\def\Tr{\text{Tr}}
\def\Vol{\text{Vol}}

\def\bk{\mathbf{k}}
\def\bx{\mathbf{x}}
\def\by{\mathbf{y}}
\def\bxi{\mathbf{\xi}}

\def\beq{\begin{eqnarray}}\def\eeq{\end{eqnarray}}
\def\be{\begin{equation}}\def\ee{\end{equation}}

\title{Entanglement Entropy in (3+1)-d Free $U(1)$ Gauge Theory}

 \author{Ronak M Soni,}
 \author{Sandip P. Trivedi}
 \affiliation{\it Department of Theoretical Physics,
 Tata Institute of Fundamental Research,\\  Colaba, Mumbai, 400005, India}

\vspace{5mm}

\vspace{1cm}

\emailAdd{ronak@theory.tifr.res.in}
\emailAdd{sandip@theory.tifr.res.in}

\abstract{We consider the entanglement entropy for a free $U(1)$ theory in $3+1$ dimensions in the extended Hilbert space definition. 
By taking the continuum limit carefully we obtain a replica trick path integral which calculates  this entanglement entropy. The  path integral is gauge invariant, with a gauge fixing delta function accompanied by a  Faddeev -Popov determinant. 
For a spherical region  it follows that the result for the logarithmic term  in the entanglement, which is universal, is given by the $a$ anomaly coefficient. 
We also consider the extractable part of the entanglement, which corresponds to the number of Bell pairs which can be obtained from entanglement distillation or dilution. For a spherical region we show that the coefficient of the logarithmic term for the extractable part  is  different from the extended Hilbert space result. We  argue  that the two results will differ in general, and this difference is accounted for  by a massless scalar living on the boundary of the region of interest. }

\preprint{\parbox{3cm}{TIFR/TH/16-29}}

\begin{document}
\maketitle
\flushbottom

\section{Introduction}

Entanglement Entropy is being increasingly recognised as an important measure of quantum  correlations in a system. 
For gauge theories the entanglement entropy turns out to be   non-trivial to define. Physically, this is because of extended excitations in the system,  like loops of electric or magnetic flux created by  Wilson and 't Hooft loop operators. More precisely, it is because the Hilbert space of physical states in a gauge theory, i.e. the set of gauge-invariant states,  does not admit a tensor product decomposition between the region of interest, the ``inside'', and the rest of the system, the ``outside''. 

Several ways to deal with this difficulty have been discussed in the literature leading to different definitions of the entanglement entropy. 
In \cite{Ghosh2015} and \cite{Aoki2015}, a definition, called the Extended Hilbert Space definition, was given for a gauge theory on a spatial lattice, by embedding the gauge-invariant states in a larger Hilbert space which now admits a tensor product product decomposition between the inside and the outside regions.  See also earlier important work in \cite{Casini2013,Buividovich2008b,Donnelly2011,Radicevic2014}.  This definition has several positive features. It is gauge-invariant, meets  the strong subadditivity condition, and can be applied to all gauge theories; Abelian and Non-Abelian, discrete and continuous,  with and without matter. 
It was also argued in \cite{Ghosh2015} that this definition  agrees with the Replica Trick definition of Entanglement Entropy, which is an alternate way to define entanglement  based on carrying out a Euclidean path integral.

In \cite{Soni2015,vanAcoleyen2015},  the extended Hilbert space definition was analysed in more detail and it was also shown  that the definition does not agree with  an operational  measure of entanglement which comes from quantum information theory, and which is related to the entanglement that can be extracted from two halves of a bipartite system  in the form of Bell pairs, or the entanglement which is available for use in the process of   entanglement dilution. 

The underlying reason for this is the extended Wilson and 't Hooft operators mentioned above. 
Without using them, and restricting only to gauge-invariant operators localised in the inside or outside regions,  only some of the total entanglement can be extracted. In fact, the  set of inside states breaks up into  sectors differing in the value the  normal component of the electric flux  takes at the boundary of the inside region.  Wilson loop operators which cross the boundary from the inside to the outside change this boundary electric flux and  are needed to connect these sectors. Without them, the different sectors act like different superselection sectors and the probability to lie in these sectors cannot be altered. This puts a limitation on how much total entanglement can be extracted. 
\cite{Soni2015,vanAcoleyen2015}   precisely determined just how much of the total entanglement entropy could actually be extracted in entanglement distillation and dilution. Unlike the total entanglement entropy which depends on the extended Hilbert space definition that is used to define it, the extractable part is tied to  an  operational definition, based on physical measurements, and is  independent of this definition.

In this paper we explore the extended Hilbert Space definition in the concrete context of a free $U(1)$ gauge theory without matter in $3+1$ dimensions. 
Our main focus here  is  the continuum limit and this is one of the simplest theories to admit such a  limit.
By starting with the theory on the lattice at weak coupling  and taking the continuum limit, we show that one obtains from this definition  the replica trick path integral for the continuum $U(1)$ theory. More specifically, we obtain the replica trick path integral with  a suitable Faddeev-Popov gauge-fixing term to render it gauge-invariant. While we focus on the $U(1)$ theory, this result is in fact general and extends to non-Abelian theories, and theories with matter etc, also. 

In $3+1$ dimensions the entanglement entropy has the behaviour 
\begin{equation}
\label{see}
S_{EE}=C_1{A\over \epsilon^2} + C \log(A/\epsilon^2) + \cdots
\end{equation}
where $A$ is the area of the region of interest, $\epsilon$ is a short distance cut-off and the ellipses denoted finite terms in which $\epsilon$ does not appear. 
The coefficient $C$ is cutoff independent and therefore universal. 

For a spatial region which is the inside of  a sphere, $S^2$,  we obtain from the replica trick path integral for the $U(1)$ theory, by a standard argument,  the coefficient $C$.
It is given by
\begin{equation}
\label{valC}
C=-{31\over 90}
\end{equation}
and  agrees with the $A$ anomaly coefficient in the theory. 

Next we analyse how much of the entanglement is extractable through the processes of entanglement distillation and dilution in this system. As was mentioned above,  sectors differing in the value  the normal component of the electric flux takes at the boundary of the inside region act as different superselection sectors. 
Let $i$ schematically denote such a  sector and let  $p_i$ be  the probability for the normal component of the boundary electric flux associated with this sector  to arise. Then  it was argued in \cite{Soni2015,vanAcoleyen2015} that for an Abelian theory, 
\begin{equation}
\label{extsee}
S_{EE}=-\sum_i p_i\log p_i + S_{ext},
\end{equation}
where $S_{EE}$ is the full entanglement in the extended Hilbert space definition and $S_{ext}$ is the amount of  entanglement which can be extracted using operators localised in the inside and outside regions in the processes of entanglement distillation and dilution. We see that the full entanglement differs from the extractable piece by a ``classical'' Shannon -like term determined by the probability distribution for the electric flux at the boundary of the inside  region. Let us also clarify that eq.(\ref{extsee}) is only schematic. In general the probability will be a functional of the normal component of the electric field which can take varying values along the boundary, and the first term in eq.(\ref{extsee}) will be a functional integral over all such values for the electric field. 

By analysing the probability distribution for the electric field on the boundary  $S^2$ region  we find that the extractable piece also has a term proportional to $\log(A)$,
\begin{equation}
\label{extract}
S_{extractable}=D_1 {A\over \epsilon^2} + D \log({A\over \epsilon^2}) + {\rm finite}
\end{equation}
 where
  \begin{equation}
 \label{valD}
 D=-{16\over 90}
 \end{equation}
 The difference $C-D$ is accounted for by the first term in eq.(\ref{extsee}) which also has a contribution proportional to  $\log(A)$. 

There has been some confusion in the literature on the coefficient of the logarithmic term for the $U(1)$ theory. 
Conformal mappings and related techniques, it has been known for some time, give, eq.(\ref{valC}) \cite{Solodukhin2008,Casini2011}. On the other hand, the calculation done by Dowker \cite{Dowker2010}  gave the result, eq.(\ref{valD}). More recently, Casini and Huerta \cite{Casini2015}, using a definition different from the extended Hilbert space definition used here, obtained the same result as Dowker also agreeing therefore with eq.(\ref{valD}). 
Our results show clearly that  these differences are due to different definitions being adopted for the entanglement entropy. The total extractable piece, eq.(\ref{extract}),  is tied to the number of Bell pairs one can obtain from the system, which is an operational quantity with physical significance.  Similarly, this difference in definitions can drop out of quantities constructed from the entanglement  like the mutual information in the continuum limit \cite{Casini2013,Casini2014}.


This paper is organised as follows. After a brief outline of the extended Hilbert space definition, we turn to the connection with the replica trick path integral in section \ref{sec:replica-trick-1} and take its continuum limit via the replica trick in section \ref{sec:replica-trick-2}. The following sections, \ref{sec:u1} and \ref{sec:plogp}, then discuss the coefficients $C$ and $D$ respectively, eq. \eqref{finvalc}, and \eqref{extracb}.
We end with some  conclusions in \ref{sec:conclusion}.

Before we close let us comment on some of the relevant literature. 
Besides the references above, some key references in the discussion of gauge theory entanglement are \cite{Kabat1995,Buividovich2008b,Donnelly2011,Casini2013,Radicevic2014,Casini2014,Donnelly2014,Donnelly2014b,Huang2014,Ghosh2015,Aoki2015,Donnelly2015,Radicevic2015,Soni2015,vanAcoleyen2015,Casini2015,Zuo2016}.
The replica trick has been discussed in many places, for example \cite{Solodukhin2011,Cardy2007}.
The calculations in section \ref{sec:plogp} where we compute the extractable piece are closely related to those by Donnelly and Wall \cite{Donnelly2014b,Donnelly2015}, Huang \cite{Huang2014} and Zuo \cite{Zuo2016}.

\section{Overview of the Extended Hilbert Space Definition} \label{sec:ehs}

Consider a  gauge  theory, associated with a gauge group $G$,  on a spatial lattice. We will work  in the Hamiltonian formulation where time is continuous.
The dynamical degrees of freedom are associated with links of the lattice. We denote a link which emanates from vertex $i$ and extends to vertex $j$ by $L_{ij}$. The degree of freedom associated with $L_{ij}$ is a group element,   $g_{ij}\in G$. In the quantum theory, this degree of freedom gives rise to  a Hilbert space, ${\cal H}_{ij}$, associated with $L_{ij}$. 
As an example, for the $U(1)$ theory of interest here a group element is specified by    an angle $\theta_{ij} \in [0,2\pi]$. The Hilbert space is then the  set of  states in quantum mechanics associated with   an angle degree of freedom, or a particle on a circle. 

The extended Hilbert space is  given by the  tensor product of ${\cal H}_{ij}$ for all links,
\begin{equation}
\label{exh}
{\cal H}= \otimes {\cal H}_{ij}.
\end{equation}
Each link Hilbert space, ${\cal H}_{ij}$  is endowed with an inner product meeting the usual positivity conditions. This in turn gives rise to an inner product on ${\cal H}$ also meeting these conditions.\footnote{In the Hamiltonian formulation we set the group elements on the links directed in  the time direction to be unity, in effect setting $A_0=0$. The resulting inner product then  has positive norm.}

Gauge transformations are defined on a vertex of the lattice and  correspond to the  action by a group element  on every link emanating from this vertex. 
Physical states must be invariant under these  gauge transformations. This follows from Gauss' laws. For the $U(1)$ theory,  in the Hilbert space ${\cal H}_{ij}$ on link $L_{ij}$, there is an operator conjugate to $\theta_{ij}$, ${\cal L}_{ij}$, which satisfies the commutation relation
\begin{equation}
\label{relc}
[\theta_{ij}, {\cal L}_{ij}]=i. 
\end{equation}
Given a vertex $V_i$ we can define the operator 
\begin{equation}
\label{gt}
G_i(\theta)= e^{i\theta\sum_{j}{\cal L}_{ij}}
\end{equation}
where the index $j$ in the sum on the RHS takes values over all links emanating from vertex $V_i$.  This operator generates gauge transformations at $V_i$  corresponding to shifting $\theta_{ij}\rightarrow \theta_{ij}+ \theta$ for all links emanating  from  $V_i$. 

Given a set of links of interest, ``the inside'',  and the rest of the links, ``the outside'', two Hilbert spaces ${\cal H}_{in}, {\cal H}_{out}$, which are subspaces of ${\cal H}$, can be defined  by taking the tensor product of the Hilbert spaces associated with the inside and outside links respectively. For example, 
\begin{equation}
\label{defhin}
{\cal H}_{in}=\otimes_{\langle ij \rangle}{\cal H}_{ij}
\end{equation}
where the links included on the RHS lie in the inside, and similarly for ${\cal H}_{out}$. 

The set of gauge-invariant states lie in a Hilbert space ${\cal H}_{ginv}$. The extended Hilbert space admits a decomposition in terms of ${\cal H}_{ginv}$ and its orthogonal complement, ${\cal H}_{ginv}^{\perp}$,
\begin{equation}
\label{oc}
{\cal H}={\cal H}_{ginv} \oplus {\cal H}_{ginv}^{\perp}.
\end{equation}

In the extended Hilbert space definition, a physical state $\ket{\psi} \in {\cal H}_{ginv}$ is first uniquely  embedded in ${\cal H}$ by requiring it to have no component in ${\cal H}_{ginv}^\perp$. 
Next, a density matrix for the inside region can   be obtained by tracing over ${\cal H}_{out}$,
\begin{equation}
\label{defrho}
\rho_{in}=\Tr_{{\cal H}_{out}} \ket{\psi} \bra{\psi}.
\end{equation}
The extended Hilbert space definition for the entanglement is  then given by the von Neumann entropy of $\rho_{in}$,
\begin{equation}
\label{ehsee}
S_{EE}=-\Tr_{{\cal H}_{in}}\rho_{in} \log(\rho_{in}).
\end{equation}

We will explore some consequences of the  entanglement entropy  defined in this way for the $U(1)$ case in the continuum limit below. 
\section{The Replica Trick for Gauge Theories on the Lattice} \label{sec:replica-trick-1}
In this section we consider a replica trick path integral which calculates the entanglement entropy in the extended Hilbert space definition for the gauge theory defined on a spatial lattice. 
Te discussion holds for any gauge theory.
For concreteness though we will focus on the case of the $U(1)$ theory, without matter, in $3+1$ dimensions. Also, we consider a rectangular lattice, allowing for the lattice spacing in the space and time directions to be unequal. 

\subsection{The $U(1)$ Theory}
The degrees of freedom in this system are associated with links. For the $U(1)$ case  there is one angular degree of freedom $\theta_{\mu\nu}$ associated with each link $L_{\mu\nu}$, with $\theta_{\mu\nu}\in [0,2\pi]$. 

 The action of this theory is given by 
 \begin{equation}
 \label{actu1}
 S={1\over g^2} \sum_{P} \left( \cos \sum _{\langle \mu\nu \rangle}\theta_{\mu\nu} -1 \right),
 \end{equation}
 where the over all sum is over all elementary plaquettes of the lattice, schematically denoted by $P$, and the sum within the argument of the cosine is over all links in the elementary  plaquette.
 $g^2$ is a dimensionless coupling constant, which plays the role of $\hbar$. We will be interested in the $U(1)$ theory at weak coupling, where $g^2\rightarrow 0$. 
 Note that the sum in eq.(\ref{actu1}) includes plaquettes with links extending along both space and time directions. 
 
 Starting from eq.(\ref{actu1}) we can go to the Hamiltonian description in which time is taken to be continuous. This is the formulation we will use in much of the discussion that follows. 
 To obtain this description, we first set the link variables for  links that extend in the time direction to  unity, by doing suitable gauge transformations. Next, we   take the continuum limit along the time direction by making the lattice space along the time direction to go to zero, while keeping the lattice spacing along the spatial directions fixed. 
  
 The dynamical variables of the system are then the link variables along each link of the spatial lattice $\theta_{ij}$. It is easy to see that the Hamiltonian for the system is
 \begin{equation}
 \label{defH}
 H=\sum_{\langle ij\rangle }g^2 {\cal L}_{ij} - {1\over g^2} \sum_P\left[\cos\sum_{\langle ij\rangle } \theta_{ij}-1\right]
 \end{equation}
 Here ${\cal L}_{ij}$ are conjugate variables of $\theta_{ij}$. We have also absorbed a factor of the lattice cutoff along the time direction  in $H$ to make it dimensionless.
 In the quantum theory ${\cal L}_{ij}, \theta_{ij}$ 
  satisfy the commutation relations
  \begin{equation}
  \label{commrel}
  [\theta_{ij},{\cal L}_{ij}]=i
  \end{equation}
  Here the subscripts $i,j$ refer to the same link variable. For different links, the  variables $\theta, L$ commute. 
  
  As was mentioned above, the Hilbert space associated with each link ${\cal H}_{ij}$ is that of an angular degree of freedom. The extended Hilbert space ${\cal H}$ is obtained by taking 
  the tensor product, eq.(\ref{exh}). 
  All physical states must be invariant under spatial gauge transformations.  If $\ket{\psi}$ is such a state, and $V_i$ is a vertex in the spatial lattice then  this implies  
  \begin{equation}
  \label{condgi}
  e^{i \epsilon_i \sum_{j}{\cal L}_{ij}} \ket{\psi}=0  \end{equation}
where the sum is over all links emanating from the vertex $V_i$, and $\epsilon_{i}$ denotes an infinitesimal parameter.
  
  Note that under the gauge transformation, eq.(\ref{condgi}),
  \begin{equation}
  \label{transt}
  \theta_{ij}\rightarrow \theta_{ij}-\epsilon_i
  \end{equation}
  for all links $\langle ij \rangle$ emanating from vertex $i$.
  We will also use the notation  $\theta_{ji}=-\theta_{ij}$ for  the angular variable on the  oppositely oriented link. Carrying out a gauge transformation with parameters $\epsilon_i, \epsilon_j$ at adjacent vertices $V_i, V_j$, we get that  $\theta_{ij}$ transforms as 
  \begin{equation}
  \label{trans}
  \theta_{ij}\rightarrow \theta_{ij}-\epsilon_i+\epsilon_j.
  \end{equation}

 \subsection{The Replica Trick}
 We now turn to the replica trick path integral.
 Suppose we are interested in the entanglement between a region $R$ and the rest  of the system for the ground state of the theory discussed above. Let us denote the three dimensional spatial lattice by $L^3$ below. 
 To calculate the entanglement entropy  using the replica trick  we first calculate a Euclidean path integral on an n-fold cover of $L^3 \times T$, where $T$ is the extent in the  continuous imaginary time direction. The entanglement entropy is then recovered by taking a suitable $n \rightarrow 1$ limit.  
 
 The $n$-fold cover is obtained as followed.  The path integral extends from $[-\infty, \infty]$ in the $T$ direction. We take $n$ copies of $L^3 \times T$. 
 In  each copy, for the $L^3$ spatial lattice, at a particular instance $t=0$,  we  introduce a cut along the spatial region $R$ of interest.  Fields in the path integral are taken to be discontinuous along the cut. Let us generically denote  a field  by $\phi_i(t, x)$, where the subscript $i$ denotes its value in the $i^{th}$ copy. Then $\phi_i$ satisfies the relation
 \begin{equation}
 \label{relphi}
 \phi_i(t=0^{-}, {\bf x}) = \phi_{i+1}(t= 0^{+} , {\bf x}), x\in R
 \end{equation}
 So that the value of the field below the cut in the $i^{th}$ copy is identified with its value above the cut  in the $(i+1)^{th}$ copy. 
 For the $n^{th}$ copy the value below the cut is identified with the value for the $1^{st}$ copy above the cut.
 Outside the region $R$ the fields are continuous across the $t=0$ surface meeting the condition,
 \begin{equation}
 \label{relphi2}
 \phi_i(t=0^{-}, {\bf x}) = \phi_{i}(t= 0^{+}, {\bf x}), x \not \in R.
 \end{equation}
 These boundary conditions define the $n$-fold cover. 
 
 Let us denote the path integral over the $n$-fold cover by $Z(n)$. The Replica trick value for the entanglement entropy is then given by 
 \begin{equation}
 \label{entrt}
 S_{RT}=-\partial_n \ln Z(n)|_{n \rightarrow 1} + \ln Z(1)
 \end{equation}
 Note, to obtain the RHS one needs to continue the function $Z(n)$ for non-integer values, and then take the derivative.
 Subtleties  may arise in this continuation, see \cite{Cardy2007} for a discussion, but we ignore them  here. 
 
 Note that the above description is general and immediately applies to the $U(1)$ theory as well. 
 In this case the  field $\phi$ corresponds to a phase $U_{ij}=e^{i\theta_{ij}}$ on each spatial link. 
 The region $R$ will be specified by a set of spatial links. And the condition eq.(\ref{relphi}) identifies the value of the phase on a spatial link in $R$ below the cut, in the $i^{th}$ copy with its value above the cut in the $(i+1)^{th}$ copy.

 It was argued in \cite{Ghosh2015} quite generally that this replica trick path integral agrees with the extended Hilbert space definition. 
 The reason is as follows.
 Consider one copy of $ L^3\times T$. The path integral from $t=[\infty, 0^{-}]$ essentially gives rise to the wave function of the ground state. If $U_{ij}(0^{-})$ is the value the link variables take 
 on link $L_{ij}$ at $t=0^-$ then on general grounds we get that 
 \begin{equation}
 \label{gengr}
 \braket{U_{ij}(0^-)}{\psi}=\int_{t=-\infty}^{U_{ij}(0^{-} )} [DU_{ij}]   e^{-S}.
 \end{equation}
 On the RHS the action which appears, $S$,  is obtained  from  the Hamiltonian by the standard relation  between    the path integral and  time evolution obtained from the Hamiltonian. The boundary conditions are that the link variables take values $U_{ij}$ at $t=0^-$, which are the arguments for the wave function on the LHS.
 The boundary conditions at $t=-\infty$ drop out, except in determining the total normalisation of $\ket{\psi}$. 
 The reason that the path integral in eq.(\ref{gengr})  gives rise to the ground state wave function is that only the lowest energy state survives after time evolving from $t=-\infty$ to $t=0^{-}$.
 
 For the $U(1)$ theory  $S$ is given by eq.(\ref{actu1}) with the  link variables along the time like links being set to vanish.
  Similarly we get that 
 \begin{equation}
 \label{brawf}
 \braket{\psi}{U_{ij}(0^+)}=\int_{U_{ij}(0^+)}^{t=\infty}D[U] e^{-S}.
 \end{equation}
 
It is now easy to see that doing the path integral from $t=-\infty$ to $t=\infty$  with discontinuous boundary conditions at $t=0$ across the region $R$ gives rise to the density matrix for the 
region $R$:
\begin{equation}
\label{densr}
\mel{U_{ij}(0^-)}{\rho}{U_{ij}(0^+)}=\int [DU_{ij}] e^{-S},
\end{equation}
where 
\begin{equation}
\label{defrho}
\rho=\ket{\psi}\bra{\psi}.
\end{equation}
In the path integral on the RHS of eq.(\ref{densr}) $U_{ij}$ takes the value $U_{ij}(0^+)$ for $t=0^+$, and $U_{ij}(0^-)$, for $t=0^-$ for links lying in $R$,
and is continuous for links outside $R$, so that 
\begin{equation}
\label{outU}
U_{ij}(0^+)=U_{ij}(0^-), \langle ij\rangle\not \in R,
\end{equation}
in agreement with  eq.(\ref{relphi2}). 

The key point for our present discussion is that this density matrix agrees with what one obtains in the extended Hilbert space definition. The reason is simply that in carrying out the path integral
each link variable $U_{ij}$ is independent of the others, and as a result the wave function one obtains in eq.(\ref{gengr})  by the path integral is the wave function automatically embedded 
in ${\cal H}$, the extended Hilbert space.  The boundary condition for links in the outside region, eq.(\ref{outU}), then means that the path integral from $t=[-\infty, \infty]$  carries out the trace over the outside links to give the density matrix of the inside,
eq.(\ref{densr}).

It then follows that the path integral over the $n$ fold cover, 
\begin{equation}
\label{defzn}
Z(n)=\Tr\rho^n,
\end{equation}
  and standard manipulations then lead to the conclusion that $S_{RT}$ defined in eq.(\ref{entrt}) is in fact the entanglement entropy $S_{EE}$ in the extended Hilbert space definition, eq.(\ref{ehsee}).

\section{The Continuum Limit of the Extended Hilbert Space Definition} \label{sec:replica-trick-2}
 In this section we will analyse the behaviour of the entanglement entropy, defined in the extended Hilbert space definition, in the  continuum limit of the gauge theory.
 Since we have shown that on the lattice the extended Hilbert space definition is equivalent to a replica trick path integral, we can study this limit by analysing the continuum limit of the  replica trick path integral.
  
  In the continuum limit the lattice spacing, $\epsilon \rightarrow 0$, keeping physical distances fixed. In particular we are interested in the behaviour of the entanglement entropy, for the ground state of the gauge theory, of  region $R$ which is kept fixed as $\epsilon \rightarrow 0$. Note also that  in the Hamiltonian formulation $\epsilon$ 
  refers to the  size of the spatial lattice, since the lattice spacing in the time direction has already been taken to vanish, as discussed at the beginning of section \ref{sec:replica-trick-1} above. 
  
  The continuum limit needs to be taken carefully. There are two complications. The first one is the standard complication  associated with taking the continuum limit in quantum field theory, and arises because the coupling constants of the theory need to be renormalised in a suitable way as $\epsilon \rightarrow 0$. We are spared this complication for the weakly coupled $U(1)$ theory under consideration since it goes over to the Maxwell theory in the continuum limit, which is free. However there is another complication which arises  for entanglement  that cannot be avoided even in the $U(1)$ case. 
 This arises because the path integral involved in calculating the  entanglement, in the continuum limit, needs to be carried out over a singular space and not a smooth manifold. For example, for computing $Z(n)$, the partition function on the $n$-fold cover,  the path integral needs to be carried out over a  singular space with a conical singularity of definite angle $2\pi(n-1)$
   along the boundary of the region $R$. The singular nature of this space gives rise to additional divergences which need to be regulated. 
   As mentioned above  the leading and sub-leading divergences in $3+1$ dimensions take the form, eq.(\ref{see}). We are in particular interested in the coefficient $C$ of the term proportional to  $\log(A)$ that arises in the continuum limit.  In the discussion below we will take  $\epsilon$ to be small but non-zero and study the resulting path integrals. 
   \subsection{$Z(1)$: The Partition Function}
   Let us start by considering $Z(1)$ first. From eq.(\ref{defzn}),  eq.(\ref{densr}) we see that $Z(1)$ is obtained by sewing the bra and ket of the ground state  wave function $\ket{\psi}, \bra{\psi}$ together,
   \begin{equation}
   \label{valz1a}
   Z(1)=\Tr\rho = \int [DU] \braket{U_{ij}}{\psi}\braket{\psi}{U_{ij}}.
   \end{equation}
   The path integral on the left hand side is over gauge fields $U_{ij}$ living on  link variables in the spatial lattice at $t=0$.
   And 
   in this case the path integral is over one copy of $L^3\times T$ since the fields are continuous at $t=0$. From eq.(\ref{densr}) we get
   \begin{equation}
   \label{respiz1}
   Z(1)=\int [DU] e^{-S},
   \end{equation}
   where the path integral on the RHS is now from $t\in[-\infty,\infty]$.
   The measure  in eq.(\ref{respiz1}) is defined as follows. We discretise the time direction as well to take values $t_i={T\over N}i$ with  $i\in [-N/2,N/2]$. 
   At each time step the link variables are denoted by $U_{ij}(t_i)$.  The measure for each link variable at a time step is the Haar measure. 
   The full measure $DU$
   is then given by the product of the measures  for link variables at each time step. In the $U(1)$ case the measure on each link is given by 
   \begin{equation}
   \label{mlink}
   [DU_{ij}]={d\theta_{ij}\over 2\pi}
   \end{equation}
   with $\theta\in[0,2\pi]$.
   This measure is invariant under the gauge transformation, eq.(\ref{trans}). 
   
   Since $\ket{\psi}$ is gauge-invariant,  
   \begin{equation}
      \label{gtinv}
      \braket{U_{ij}}{\psi} = \braket{U_{ij}^{g}}{\psi},
      \end{equation}
      where $U_{ij}^g$ is a transformed value of the link variables obtained after a gauge transformation  schematically  denoted  by superscript $g$ here.  For the $U(1)$ theory this is of the type given in eq.(\ref{trans}).  
      
      As a result  the path integral over the gauge fields at $t=0$ in eq.(\ref{valz1a}) has a redundancy since $\braket{U_{ij}}{\psi} \braket{\psi}{U_{ij}}$  yields the same result  for different values of the link variables $U_{ij}$ related by gauge transformations. This redundancy is not a problem in the lattice since the integration is over a group manifold which is compact.  In the $U(1)$ case this is simply the fact that $\theta_{ij}$ is periodic. In the continuum limit however, as we see below, we shall  replace the angular degrees of freedom by the gauge potential $A_i$ which is  taken to be non-compact. 
Before taking this limit  it is therefore important to make this gauge  redundancy manifest. This can be  done in the standard fashion by breaking  up the path integral in eq.(\ref{respiz1}) into two parts: a sum over gauge-inequivalent configurations and for each such choice a further sum over all gauge-transformed values of these configurations.

    Let $\phi({\bf n})$ denote the  parameter for a gauge transformation at site ${\bf n}$. Then it is a mathematical identity that 
      \begin{equation}
      \label{maid}
      \int \left[\prod_{\bf n\in L^3} d\phi({\bf n})\right]  \   \delta\left(f(\phi_{\bf n})\right)   \ \left|\det \left({\partial f\over \partial {\phi_{\bf n}}}\right)\right| =1.
      \end{equation}
      Here the product is over all  $N_V$ sites of the spatial lattice and $f(\phi_{\bf n})$ is actually condensed notation for  a set of functions $f_i, i, \cdots N_V$ such that the conditions $f_i(\phi_{\bf n})=0, $  determines  the  gauge parameters, $\phi_{\bf n}$, on all sites uniquely.  Similarly, the determinant on the RHS is for the $N_V\times N_V$ matrix, ${\partial f_i\over \partial {\phi_{\bf n}}}$.

      Introducing this identity in eq.(\ref{valz1a}) and rearranging terms gives
      \begin{equation}
      \label{revaz1}
      Z(1)=\int [D\phi] \int [DU_{ij}]  \ |\psi(U_{ij})|^2  \ \delta(f(\phi_{\bf n})) \left|\det \left({\partial f\over \partial {\phi_{\bf n}}}\right)\right|,
      \end{equation}
      where $[D \phi]$ denotes the measure $\prod_{\bf n} d\phi({\bf n})$.
      
      Next we take $f(\phi_{\bf n})=f(U_{ij}(\phi_{\bf n} ))$, i.e. $f$ to depend on the gauge transformation parameters $\phi({ \bf n})$ only implicitly through its  dependence on $U_{ij}$.      Using the invariance of the measure  under a gauge transformation, as discussed above,  and also the invariance of the wave function, we then get 
      \begin{align}
            \int DU_{ij} \ |\psi(U_{ij})|^2 \  \delta(f(\phi_n)) \left|\det \left({\partial f\over \partial {\phi_{\bf n}}}\right)\right| &= \int DU_{ij}(\phi_n)|\psi(U_{ij}(\phi_n))|^2 \delta(f(\phi_n)) \left|\det \left({\partial f\over \partial {\phi_{\bf n}}}\right)\right| \label{invpia}\\
      &=  \int DU_{ij}|\psi(U_{ij})|^2 \delta(f(U_{ij}) \left|\det \left({\partial f\over \partial {\phi_{\bf n}}}\right)\right| \label{invpib}
      \end{align}
      Note, in the first line on the RHS, $U_{ij}(\phi_{\bf n})$ denotes the element of the group obtained from $U_{ij}$ after the gauge transformation generated by $\phi_{\bf n}$.
           The second line is  then obtained by relabelling the integration variable of the RHS in the first line, $U_{ij}(\phi_n)$, to be  $U_{ij}$. 
      On general grounds one can argue that the determinant $\det\left({\partial f\over \partial {\phi_n}}\right) $ is independent of $\phi_n$. Using eq.(\ref{invpib})   in eq.(\ref{revaz1})
       then gives
      \begin{equation}
      \label{revaz2}
      Z(1)=\int [D\phi] \ \left( \int DU_{ij}|\psi(U_{ij})|^2 \delta(f(U_{ij})) \left|\det \left({\partial f\over \partial \phi_{\bf n}}\right)\right| \right),
      \end{equation} 
      where the terms within the big brackets are independent of $\phi_n$. 
      
      The integral over $\phi_n$ can  therefore  be easily done. For the $U(1)$ case we get 
      \begin{equation}
      \label{iphi}
      \int [D\phi]=\prod_n\int_0^{2 \pi}  d \phi_n =(2\pi)^{N_V},
      \end{equation}
      where $N_V$ is the total number of sites on the spatial lattice. 
      For a more general gauge group we would get $(\Vol(G))^{N_V}$ where $\Vol(G)$ is the volume of the group manifold as computed from the Haar measure. 
      Pugging eq.(\ref{iphi}) in eq.(\ref{revaz2})  then gives, 
      \begin{equation}
      \label{revaz3}
      Z(1)=  (2\pi)^{N_V} \int DU_{ij}|\psi(U_{ij})|^2 \delta(f(U_{ij})) \left|\det \left({\partial f\over \partial \phi_{\bf n}}\right)\right|.
      \end{equation}
      
      One choice of the set of functions $f_{\bf n}$ is given as follows. A vertex denoted by ${\bf n}$ has six nearest neighbours in the spatial lattice, let us denote them  by ${\bf n\pm {\hat i} }$, where 
      ${\bf {\hat i}},  {\bf {\hat i}} ={\bf  {\hat 1},{\hat 2},{\hat 3}}$, stand for  the lattice  unit vectors along the $x,y,z$ directions.  The link variable on the link extending from ${\bf n} $ to 
      ${\bf n+ {\hat i}}$ is denoted by $\theta_{{\bf n}, {\bf n+{\hat i}}}$ etc. 
      We take  
      \begin{equation}
      \label{chf}
      f_{\bf n}=\sum_{\bf{\hat  i}} \theta_{{\bf n}, {\bf n+ {\hat i}}}+\theta_{{\bf n}, {\bf n-{\hat i}} }.
      \end{equation}
      Under a gauge transformation generated by the set $\{\phi_{n}\}$ this transforms as 
      \begin{equation}
      \label{chf2}
      f_{\bf n}\rightarrow f_{\bf n}+ \sum_{\bf{\hat  i}}(2\phi_{\bf n}-\phi_{\bf n + {\hat i}}-\phi_{\bf n-{\hat i}}).
      \end{equation}
      It is easy to see that the conditions $f_{\bf n}=0, \forall  {\bf n} \in L^3$, fixes all the gauge redundancy\footnote{These conditions fix $\phi_{\bf n}$ upto  one overall gauge transformation, but the link variables do not transform under this  transformation, so it is not a redundancy of the variables in the  gauge theory.} 
     and the  resulting determinant $\det \left({\partial f_{\bf n}\over \partial \phi_{\bf m}}\right)$ is independent of $\{\phi_{\bf n}\}$.
      
       We can now take the continuum limit of the integral which appears on the RHS of eq.(\ref{revaz3}).    
        We assume that the path integral is dominated by configurations which are smooth on the scale of the lattice, we will come back to discussing this assumption below. 
   The $\cos(\sum_{<ij>}\theta_{ij})$ term in the action, eq.(\ref{actu1}) then can be expanded unto quadratic order. Consider a plaquette extending in the $i-j$ directions. For this plaquette we get
    \begin{equation}
   \label{cosexp}
   \cos(\sum_{<ij>}\theta_{ij})\simeq 1-{1\over 2} (\theta_{{\bf n,  n + {\hat i} }} -\theta_{{\bf n +{\hat j}, n +   {\hat i} + {\hat j} }}+ \theta_{\bf{n + {\hat i}, n+{\hat i}+{\hat j}}}-
   \theta_{\bf{n,n+{\hat j} }} )^2.
   \end{equation}
   The kinetic energy term, from eq.(\ref{actu1}),  is given by 
   \begin{equation}
   \label{KEdis}
   KE= {\epsilon^2\over 4g^2}  \sum_{{\bf n, {\hat i}}} ({\dot \theta}_{\bf n, n+{\hat i}})^2.
   \end{equation}
   where the prefactor of  ${1\over 4}$ is because each link is being counted twice, and $\epsilon$ is the lattice cut-off\footnote{We are not distinguishing between the cut-off in the time and spatial directions here.}.
   
   It can be then easily seen that the  wave function eq.(\ref{gengr}) goes over to the   continuum expression
   \begin{equation}
   \label{conwf}
   \braket{A_i({\bf x}, 0^-)}{\psi}=\int_{t=-\infty}^{A_i({\bf x}, 0^{-})} [DA] e^{-S_M} ,
   \end{equation}
   with the identifications
   \begin{equation}
   \label{valA}
   A_i({\bf x, t}) ={1\over g \epsilon} \theta_{\bf n, n+{\hat i}}
   \end{equation}
   and the action $S$  which appears in eq.(\ref{conwf}) is the Maxwell action
   \begin{equation}
   \label{maxwell}
   S_M={1\over 4 } \int d^4 x F_{\mu\nu}F^{\mu\nu}
   \end{equation}
   with $A_0=0$. 
   The path integral, eq.(\ref{conwf}), is carried out over non-compact variables $A_i$ with the standard continuum measure. 
   
   Note the fact that the spatial gauge transformations are unfixed in path integral in eq.(\ref{conwf})  does not pose a problem since specifying the value of $A_i({\bf x}, 0^-)$ breaks this symmetry.    However it is important that we fixed this residual gauge symmetry on  the lattice for the partition function $Z(1)$, eq.(\ref{revaz3}), otherwise we would have got a 
   divergent answer. 
   From eq.(\ref{revaz3})  we now get, 
   \begin{equation}
   \label{contz1}
   Z(1)=(2\pi)^{N_V} \int [DA_i] |\psi[A_i]|^2 \delta(f(A_i)) \left|\det \left({\delta f(A^\omega_i)\over \delta w({\bf x})}\right)\right|,
   \end{equation}
   where $\omega(x)$ is the gauge transformation parameter as the location ${\bf x}$, and $A^\omega_i$ is the gauge potential obtained after gauge transforming $A_i$ under this gauge transformation.  
   
   For the choice eq.(\ref{chf}) we get
   \begin{align}
   f(A_i) & = \nabla \cdot A \label{conaa}\\
   \det \left({\delta f(A^\omega_i)\over \delta w({\bf x})}\right)& = \det\nolimits' (\nabla^2) \label{conab}
   \end{align}
   The prime on the determinant on the RHS of the second line indicates that the zero mode has been removed (this mode does not change $A_i$ and  does  not need to be included). 
   
   Using eq.(\ref{conwf})  and the analogous expression for $\braket{\psi}{A_i}$ we can write eq.(\ref{contz1}) as 
   \begin{equation}
   \label{contz2}
   Z(1)=(2\pi)^{N_V}\int [DA_i] e^{-S_M} \delta(f(A_i)) \left|\det \left({\delta f(A^\omega_i)\over \delta w({\bf x})}\right)\right|
   \end{equation}
   We should clarify that on the RHS  the path integral in eq.(\ref{contz2}) is now from $t\in [-\infty,\infty]$, but the delta function fixing the spatial gauge transformations and associated determinant is only present at $t=0$. 
   
   To obtain the more conventional result of the Path integral with gauge-fixing at all times  we return to the expression for path integral on the lattice in eq.(\ref{revaz3}),  using eq.(\ref{gengr}) and eq.(\ref{brawf}) this can be written as 
   \begin{equation}
   \label{rez1l}
   Z(1)=(2\pi)^{N_V} \int D[U_{ij}]e^{-S} \delta(f(U_{ij})) \left|\det \left({\partial f\over \partial {\phi_n}}\right)\right|
   \end{equation}
   where again the path integral is from $t\in[-\infty,\infty]$ but the delta function is only at $t=0$.
   
   As was mentioned above,  the path integral should be thought of more correctly by breaking up the time direction  also into discrete time steps, 
   $t_i, i=[-N/2, N/2]$. For any value of $t_i$ other than $t=0$, we can  introduce gauge parameters $\phi_{\bf n}(t_i)$ for independent gauge transformation at  sites in the spatial lattice at 
   this time step using eq.(\ref{maid}). This introduces the gauge-fixing delta function and FP determinant now at the time step $t_i$ along with the extra 
   integrals for $\phi_{\bf n}(t_i)$. 
   The parameter $\phi_{\bf n}(t_i)-\phi_{\bf n}(i+1)$ can be associated with the link variable along the time direction going from site ${\bf n}$ at $t_i$ to $ {\bf n}$ at $t=i+1$.
   Note that since $\phi_{\bf n}(t=0)=0$, the counting is just right, there are as many time like links as parameters $\phi_{\bf n}(t_i)$. 
   
   After some further manipulations analogous to the change of variables in eq.(\ref{invpia}), eq.(\ref{invpib}), we then get in the continuum limit,    
   \begin{equation}
   \label{rez2l}
   Z(1)=(2\pi)^{N_V}\int [DA_i] [DA_0] e^{-S_M} \delta(f(A_i)) \left|\det \left({\delta f(A^\omega_i)\over \delta w({\bf x})}\right)\right|
   \end{equation}
   Now the  integral is over both $A_i$ and $A_0$. The action $S$ which appears in eq.(\ref{rez2l}) is the Maxwell action, including the contribution from $A_0$. And the gauge-fixing delta function and associated Faddeev-Popov  (FP) determinant are  present at each time step.
   Eq.(\ref{rez2l}) is the standard path integral for the $U(1)$ gauge theory after FP gauge-fixing.  E.g., for the choice, eq.(\ref{conaa}),  eq.(\ref{conab}) this gives the  usual path integral in Coulomb gauge.
   
   Note that the path integral on the RHS in eq.(\ref{rez2l}) is indeed gauge-invariant. The usual arguments tell us that any dependence on the choice of function $f$, which fixes gauge, drops out when we include the FP determinant, The fact that we obtained this gauge-invariant form of the continuum path integral is expected since it has been argued on general grounds, \cite{Ghosh2015}, that the extended Hilbert space definition gives a gauge-invariant result.  The   careful manipulations also  yielded the extra prefactor of $(2\pi)^{N_v}$.

   Two comments are now in order. 
   First, we have assumed above that the path integral is dominated by configurations which are smooth on the scale of the lattice. 
   In the classical theory, which follows from taking the $g^2\rightarrow 0$ limit, the ground state is 
    one where $\cos(\sum_{<ij>}\theta_{ij})=1$ for all plaquettes. 
    This condition can be met by setting 
    \begin{equation}
    \label{condgsa}
    \theta_{ij}=0,
    \end{equation}
   up to gauge transformations. 
    In the quantum theory there will be zero point fluctuations due to the uncertainty principle, and $\theta_{ij}$ will acquire a spread. At weak coupling these fluctuations will be suppressed.     
    
    This can be easily estimated. Consider a mode with spatial momentum $k\ll 1/a$ where, $a$ is the lattice spacing in the spatial directions. For such a mode it is easy to see that dispersion relation takes the form 
    \begin{equation}
    \label{disrel}\omega^2 \simeq k^2
    \end{equation}
    where $\omega$ is the frequency of the mode.
    And the spread is given by 
    \begin{equation}
    \label{condb}
   \langle (\theta_{ij})^2\rangle \sim g^2 {a \over k}
   \end{equation}
   instead of eq.(\ref{condgsa}). 
   We see that as $g^2 \rightarrow 0$ the spread also vanishes. 

Thus, at weak coupling fluctuations are suppressed and we expect that smoothly varying configurations will dominating justifying the expansion unto quadratic order in eq.(\ref{cosexp}). 

Second, the fact that the time evolution from $[-\infty, 0]$ gives the ground state wave function, as mentioned above, eq.(\ref{gengr}), is certainly true for a system with a gap. 
In our case though, since the  spatial gauge transformations are not fixed, things are a bit more subtle. While the ground state is certainly gauge-invariant,   the spectrum of the Hamiltonian, eq.(\ref{defH}),  is un-gapped with low lying states, in the  $g^2 \rightarrow 0$ limit, which are not gauge-invariant. These modes might be present even after evolving from $[-\infty,0]$ and could contaminate the ground state wave function. However, since the potential energy term is invariant under these gauge transformations, the Hamiltonian which governs these pure gauge modes is free, and as a result the dependence of the wave function on these modes is a pure phase, 
which drops out of $|\psi|^2$. This renders the arguments  above which follow from eq.(\ref{gengr}), etc,  valid. 

\subsection{ The $n$-Fold Cover : $Z(n)$}
Having dealt with $Z(1)$ quite carefully we are now ready to consider the more general case of $Z(n)$, the partition function on the $n$-fold cover.
Actually, the $Z(2)$ case reveals all the additional points which must be dealt with so we focus on this case. Th generalisation to $Z(n)$ will then be straightforward. 

As discussed above for $Z(2)$ we start with the double cover of $L^3 \times T$. The fields at $t=0$ are discontinuous meeting boundary conditions, eq.(\ref{relphi}), eq.(\ref{relphi2}).
We will first set up the path integral on the lattice, then  write it in a form where the gauge redundancy has been made manifest by fixing the residual gauge freedom to do spatial gauge transformations at $t=0$. The resulting form of the result will then admit a well defined continuum limit. 

We denote the value of the fields at $t=0^{\pm}$ inside and outside $R$, the region of interest, on the first and second sheet by $U_{1 }^{t=0^{\pm},in}, 
U_{1}^{t=0^{\pm},out}, 
U_{2}^{t=0^{\pm}, in}, U_{2}^{t=0^{\pm},out} $, respectively.  Here we have suppressed the indices  $i,j$ which appear as  subscripts in the link variables $U_{ij}$ and  specify the precise link we are referring to. Then 
the boundary conditions used for sewing up the path integral are
\begin{align}
U_{1}^{t=0^{+}, out}& =  U_{1}^{t=0^-,out} \equiv U_1^{out}\\
U_{2}^{t=0^{+}, out}& =  U_{2}^{t=0^-,out}  \equiv U_2^{out}\\
U_{1}^{t=0^-,in} & =  U_{2}^{t=0^+,in} \equiv U_{1}^{in}\\
U_{1}^{t=0^+,in} & =  U_{2}^{t=0^-,in} \equiv U_2^{in}.
\end{align}

This gives 
\begin{equation}
\label{resz2a}
Z(2)=\Tr\rho^2 = \int D[U_{1}^{in}] D[U_1^{out}] D[U_2^{in}] D[U_{2}^{out}] \braket{\psi}{U_2^{in}U_1^{out}} \braket{U_1^{in}U_{1}^{out}}{\psi} \braket{\psi}{U_1^{in}U_2^{out}}
\braket{U_2^{in}U_{2}^{out}}{\psi}
\end{equation}
where $\ket{U_1^{in}U_1^{out}}$ is a state which is the eigenvector of the link operators ${\hat U}^{in}\otimes {\hat U}^{out}$ with eigenvalue $U_1^{in}U_1^{out}$, etc.
Once again the indices $i,j,$ on link variables $U$ have been suppressed to save clutter. 

We now come to making the redundancy present in eq.(\ref{resz2a}), due to spatial gauge transformation at $t=0$, more explicit.. 
There are three kinds of vertices on the spatial lattices at $t=0$ in the double cover. 
Those which are in the outside region, the  inside region and on the boundary. The outside and inside vertices are those in which all links ending on the vertex lie in region outside or inside $R$. The boundary vertices are those for which some links terminating on the vertex lie inside and some outside the region $R$. 
Let us introduce a  set of delta function $f(U_{ij})$, which fix all the gauge redundancy, analogous to the $Z(1)$ case, by using the identity, eq.(\ref{maid}), now for the double cover. The steps from eq.(\ref{maid}) to eq.(\ref{revaz3}) can now be repeated for the $Z(2)$ case. 
This leads to
\begin{equation}
\label{resz2b}
Z(2)=(2\pi)^{2 N_V-N_B} \int D[U_{ij}]\braket{\psi}{U_2^{in}U_1^{out}} \braket{U_1^{in}U_{1}^{out}}{\psi} \braket{\psi}{U_1^{in}U_2^{out}}
\braket{U_2^{in}U_{2}^{out}}{\psi} \delta(f(U_{ij})) \left|\det\left({\partial f\over \partial {\phi_n}}\right)\right|
\end{equation}
The measure $D[U_{ij}]$ stands for the integration over $U_1^{in}, U_1^{out}, U_2^{in}, U_2^{out}$, the delta functions denoted schematically by $f(U_{ij})$ fix the gauge freedom at all vertices, and the accompanying determinant is the standard Faddeev-Popov determinant.
The pre factor is non-trivial and needs some explanation.  Gauge transformations can be carried out independently on the inside and outside vertices of both copies of $L^3$.
However the gauge transformations on the boundary vertices must be the same on the two $L^3$'s in order to be an invariance of the integrand in the path integral in eq.(\ref{resz2a}). For suppose $g$ denotes such a gauge transformation in a boundary vertex of the first $L^3$, then both the inside and outside links in $U_1^{in}, U_1^{out}$ which end on this vertex transform under it,
\begin{equation}
\label{transU}
(U_1^{in}, U_1^{out})\rightarrow (U_1^{g,in}, U_1^{g,out}).
\end{equation}
Now the wave function $\braket{U_1^{in} U_1^{out}}{\psi}$ is invariant under this transformation. However the term $\braket{\psi}{U_2^{in}U_1^{out}}$  is not invariant unless $U_2^{in}$ is also transformed, similarly invariance of $\braket{\psi}{U_1^{in}U_2^{out}}$ requires $U_2^{out}$ to also transform. Thus the spatial gauge symmetry for boundary vertices involves simultaneous transformation on both copies  $L^3\times L^3$.  Another way to say this is that the boundary vertices have enhanced coordination number. The requirement of  the  simultaneous transformation results in the prefactor in eq.(\ref{resz2b}) where, $N_V$ is the number of vertices in one copy of $L^3$ and  $N_B$ is the number of vertices on the boundary of the region $R$. 
More generally the prefactor would be $\Vol(G)^{2N_V-N_B}$. 

 The continuum limit  of eq.(\ref{resz2b}) can now be taken in a straightforward manner following the analogous steps from eq.(\ref{revaz3}) to eq.(\ref{rez2l}). We get 
 \begin{equation}
 \label{resz2c}
 Z(2)=(2\pi)^{(2 N_V-N_B)} \int D[A_0]D[A_i] e^{-S_{M}} \delta(f(A_i)) \left|\det\left({\partial f(A^\omega)\over \partial \omega(\bf{x})}\right)\right|,
 \end{equation}
 where now the path integral is being done on the double cover, the variable $A_0$ has been introduced along with the delta function fixing gauge and the Faddeev-Popov determinant, at each time step. 
 We note that the result above is also gauge-invariant, and independent of the choice of function $f$ made for gauge-fixing.
 Let us also comment that, as mentioned before, in the continuum limit the path integral has divergences since  the double cover is  singular with a conical deficit along the boundary of $R$. Thus, we need to carry out the path integral with the cut-off $\epsilon$ being held non-zero but small to get a well defined result. 
 
 The generalisation for the $Z(n)$ case is now immediate. Analogous arguments give
 \begin{equation}
 \label{fzn}
 Z(n)=(2\pi)^{( nN_V-(n-1)N_B)}  Z_{cont}(n)
 \end{equation}
 where $Z_{cont}(n)$ stands for the discretised version, with small $\epsilon$, of the continuum path integral
 \begin{equation}
 \label{contz}
 Z_{cont}(n)=\int D[A_0]D[A_i] e^{-S_{M}} \delta(f(A_i)) \left|\det\left({\partial f(A_i^\omega)\over \partial \omega(\bf{x})}\right)\right|.
 \end{equation}
 This path integral is over the $n$-fold cover which has a conical deficit $(n-1) 2\pi$ along the boundary of $R$. 
 
 Also, since the path integral with the Faddeev-Popov determinant is well known to be independent of the gauge-fixing function, we can choose a more general function $f$ too which depends on $A_\mu=(A_i, A_0)$,
 giving  more generally,
 \begin{equation}
 \label{contza}
 Z_{cont}(n)=\int D[A_0]D[A_i] e^{-S_{M}} \delta(f(A_\mu)) \left|\det\left({\partial f(A_\mu^\omega)\over \partial \omega(\bf{x})}\right)\right|.
 \end{equation}
 \subsection{Final Result}
 We can now calculate the entanglement entropy, using eq.(\ref{entrt}). 
 We get that 
 \begin{equation}
 \label{seea}
 S_{EE}=-N_B \log(2\pi)+ S_{EE, cont}
 \end{equation}
 where 
 \begin{equation}
 \label{seeb}
 S_{EE,cont}=-\partial_n \log Z_{cont}(n)|_{n\rightarrow 1}+ \log Z_{cont}(1).
 \end{equation}
 Since $N_B=A/\epsilon^2$ where $A$ is the area of the boundary and $\epsilon$ is the lattice cutoff, we see that the first term on the RHS of eq.(\ref{seea}) only contributes to the area law divergence of the entanglement. We will ignore this term below and study the contribution  of eq.(\ref{seeb}) to  the $\log(A)$ term in the entanglement. 
 For a more general group the factor of $(2\pi) $ in the first term in eq.(\ref{seea}) will be replaced by $\Vol(G)$, and this term will continue to only contribute to the area law divergence.

 \section{The Continuum Limit of the $U(1)$ Theory} \label{sec:u1}
  In this section we turn to the explicit calculation of interest.
  We would like to calculate   the coefficient of the $\log$ term in the entanglement, $C$, eq.(\ref{see}), for a spherical region of radius $R$.  The boundary of this region is an $S^2$ boundary with area
  \begin{equation}
  \label{vala}
  A=4\pi R^2.
  \end{equation}
   Also, in this section  we will drop the suffix ``$cont$'' when referring to the partition function $Z(n)$ or Entanglement, in the continuum limit, 
  see, eq.(\ref{contz}), eq.(\ref{seeb}) etc. 

  As is well known, the answer  for $C$ can be readily extracted from well known facts about the $U(1)$ theory since it is conformally invariant. 
  
  The argument is as follows. Consider  infinitesimally rescaling the radius of the sphere 
  \begin{equation}
  \label{radre}
  R \rightarrow R(1+\delta),
  \end{equation}
   $\delta \ll 1$,  while keeping the UV cut-off $\epsilon$ in eq.(\ref{see}) fixed. Then the change in $S_{EE}$ 
  is given by 
  \begin{equation}
  \label{chngs}
  \left. { \partial S_{EE}\over \partial \delta} \right|_{\delta = 0} = 2 C_1 {A\over \epsilon^2} + 2 C.
  \end{equation}
  The coefficient of interest, $C$, is the term on the RHS which is independent of $\epsilon$.
   Note that by scale-invariance the terms in the ellipses in eq.(\ref{see}) must be independent of $A$ and therefore do not contribute to the RHS of eq.(\ref{chngs}).  
  
The rescaling of the radius $R$ can be carried out  by rescaling the metric. 
 To analyse the consequences, consider the path integral, eq.(\ref{contza}), but now in the presence of a background metric,
 \begin{equation}
 \label{pibm}
Z[g_{\mu\nu}] =\int [DA]e^{-S[g_{\mu\nu}, A_\mu]} \delta(f(A_\mu))
\left|\det \left( {\partial f(A_\mu^{\omega}) \over \partial \omega(\bf{x})} \right)\right|.
\end{equation}
The metric appears in the action as shown explicitly above, but also in the measure and in general in the gauge-fixing delta function and associated determinant.  The stress energy tensor is given by  
\begin{equation}
\label{stresst}
\langle T_{\mu\nu}\rangle ={\delta \ln Z[g_{\mu\nu}]\over \delta g_{\mu\nu}}.
\end{equation}

For a conformal theory in $3+1$ dimensions, it is well known that 
\begin{equation}
\label{cfan}
\int \sqrt{g} \langle T^\mu_\mu\rangle =a E_4 + c W^2
\end{equation}
where $E_4$ is the integral of the Euler density, 
\begin{equation}
\label{defe4}
E_{4} = \frac{1}{64 \pi^{2}} \int \sqrt{g} \left( R_{\mu\nu\alpha\beta} R^{\mu\nu\alpha\beta} - 4 R_{\mu\nu} R^{\mu\nu} + R^{2} \right)
\end{equation}
and $W^2$ is the integral of the square of the Weyl tensor given by 
\begin{equation}
\label{defw2}
W^{2} = - \frac{1}{64 \pi^{2}} \int \sqrt{g}\left( R_{\mu\nu\alpha\beta} R^{\mu\nu\alpha\beta} - 2 R_{\mu\nu} R^{\mu\nu} + \frac{1}{3} R^{2} \right).
\end{equation}
The coefficient of the $E_4$ term is called the ``a-anomaly" coefficient. 

Now if the metric is rescaled by $g_{\mu\nu}\rightarrow g_{\mu\nu} (1+2 \delta)$ then this will accomplish the required scaling of $R$, eq.(\ref{radre}). 
The change $\partial_\delta \ln Z[g_{\mu\nu}]$ is given by, 
\begin{align}
  \partial_\delta \ln Z[g_{\mu\nu}] |_{\delta\to 0} & =  2 \int_{M} \sqrt{g} T_\mu^\mu \label{chngz} \\
&= 2 a E_4 + 2 c W^2. \label{chngz2}
\end{align}

From eq.(\ref{seeb}) we therefore get that 
\begin{equation}
\label{keye}
\left. {\partial S_{EE} \over \partial \delta}\right|_{\delta \rightarrow 0}=(1-\partial_n) \partial_\delta \ln Z[n, \delta]|_{\delta \rightarrow 0, n \rightarrow 1}
\end{equation}
Comparing with eq.(\ref{chngs}) we see that $C$ can be obtained once the RHS in eq.(\ref{keye}) can be calculated.

An important point, already emphasised, is that the $n$-fold cover is a singular space. So the strategy we can use, as in \cite{Fursaev2013}, is to first slightly ``de-singularise'' the space and then take the singular limit of interest. 
For calculating $Z(n)$  for the entanglement of the sphere of radius $R$   we work on a smoothed out space with metric, 
\begin{equation}
\label{smmet}
ds^2=r^2 d\tau^2 + {r^2 + b^2 n^2\over r^2 + b^2}dr^2 + (R+r^n c^{1-n} \cos\tau)^2 (d\theta^2 + \sin^2\theta d\phi^2)
\end{equation}
Here $b,c$ are extra parameters introduced to smooth out the conical singularity at $r=0$ along an $S^2$ of radius $R$. 
One then calculates the various integrals involved (for integer $n$) and  takes the $n \rightarrow 1$ limit,  in which the dependence on these extra parameters drops out. 

In \cite{Fursaev2013} it was shown that for the case, eq.(\ref{smmet}) one gets that 
\begin{align}
  \int \sqrt{g} R^{2} &= O \left( (n-1)^{2} \right) \nonumber\\
  \int \sqrt{g} R_{\mu\nu} R^{\mu\nu} &= 32 \pi^{2} (n-1) \nonumber\\
  \int \sqrt{g} R_{\mu\nu\alpha\beta} R^{\mu\nu\alpha\beta} &= 64\pi^{2} (n-1).
\end{align}
As a result $W^2\sim O((n-1)^2)$, while 
\begin{equation}
\label{vale4}
E_4=-(n-1).
\end{equation}
From eq.(\ref{keye}) it then follows that 
\begin{equation}
\label{valc}
C=a.
\end{equation}

In the $U(1)$ it is well known that $a=-{31 \over 90}$ leading to 
\begin{equation}
\label{finvalc}
C=-{31\over 90},
\end{equation}
which agrees with eq.(\ref{valC}).

\section{Extractable Part of Entanglement For the $U(1)$ Case} \label{sec:plogp}
As was mentioned in the introduction the entanglement entropy in the extended Hilbert space definition does not agree with the entanglement which can be extracted using entanglement distillation of dilution, $S_{ext}$. 
The relation between the two is given in eq.(\ref{extsee}). 
We have calculated the coefficient $C$ which appears in coefficient of the $\log(A/\epsilon^{2})$ term in  $S_{EE}$, eq.(\ref{see}),  above. Here we will calculate the coefficient of the log 
 term in $S_{ext}$. 
We will see that it is different. This difference will also allow us to understand   some puzzles  in the existing  discussion of the entanglement entropy  for the $U(1)$ theory. 
This calculation has been previously performed, using different techniques, in \cite{Donnelly2014b,Donnelly2015,Huang2014,Zuo2016}.

Our strategy will be to calculate the ``classical'' term, $-\sum_ip_i\log(p_i)$, which arises due to    the probability for being in different superselection sectors. The difference between $S_{EE}$ and this term will then give the extractable part, eq.(\ref{extsee}). 

The different superselection sections correspond to different values for the normal component of the electric field. For the case of interest the boundary is an $S^2$ and the normal component is the radial component of the electric field. Since the $U(1)$ theory is free the probability distribution governing the radial electric field $E_r$ on the boundary $S^2$  is  a Gaussian which is entirely determined by the two-point function. We get
\begin{equation}
\label{pe}
p[E_r({\bf x})]=N e^{-{1\over 2} \int d^2x d^2y E_r({\bf x}) E_r({\bf y}) G_{rr}^{-1} ({\bf x}-{\bf y})}.
\end{equation}
$G_{rr}({\bf x}-{\bf y})$ is the two-point function
\begin{equation}
\label{deftwopt}
G_{rr}({\bf x}-{\bf y})=\langle E_r({\bf x}) E_r({\bf y}\rangle
\end{equation}
on the sphere, and $G_{rr}^{-1}$ is its inverse which satisfies the condition, 
\begin{equation}
\label{invg}
\int d^2y G_{rr}({\bf x}-{\bf y}) G^{-1}_{rr}({\bf y} - {\bf z} )=\delta^{2}({\bf x}-{\bf z}).
\end{equation}
The integrals in eq.(\ref{pe}) and eq.(\ref{invg}) are in $2$ dimensions over the surface of the $S^2$.

It is  easy to see from eq.(\ref{pe})  that the required classical term is given by 
\begin{equation}
\label{ret}
-\sum_ip_i \log(p_i) = -\log N + \int d^2x \ d^2y \ G_{rr}({\bf x}-{\bf y}) G_{rr}^{-1}({\bf y}-{\bf x})
\end{equation}
From eq.(\ref{invg}) we see that the second term on the RHS gives
\begin{equation}
\label{std}
\int d^2x \  d^2 y G_{rr}({\bf x}-{\bf y} ) G_{rr}^{-1}({\bf y}-{\bf x})= \int d^2x \  \delta^2 (0)\sim {A \over \epsilon^2}
\end{equation}
where $A=4\pi R^2$ is the area of the $S^2$ and $\epsilon$ is the short distance cut-off which regulates the divergence in the two dimensional delta function, $\delta^2(0)$. 

We see that the $ \log(A)$ term which is our focus here will arise therefore from the first term on the RHS in eq.(\ref{ret}), $-\log N$. 
From eq.(\ref{pe}) we see that the normalisation $N$ is determined by requiring that 
\begin{equation}
\label{condpa}
\int D[E_r] p[E_r({\bf x})] = 1.
\end{equation}
 The Gaussian integral in eq.(\ref{condpa}) can be done and gives
 \begin{equation}
 \label{constn}
 \log N = D_4+ {1\over 2} \log \det G^{-1}_{rr}.
 \end{equation}
 Determining the constant $D_4$ (which actually turns out also to  diverge as $A/\epsilon^2$) requires a careful definition of the measure in the functional integral for $E_r$. Starting from the lattice and passing to the continuum gives rise to a well defined measure and thus to a normalisation constant. However, let us not be very  specific about this for now,
 since the resulting contribution does not give rise to a term proportional to  $\log(A)$ term.  We will comment on this again   towards the latter part of this section. Also, we have not kept track of possible zero modes in eq.(\ref{constn}), we will be more precise in the discussion below in this regard as well. 
 
 There is one more subtlety which  however must be addressed at the outset.  The Green's function eq.(\ref{deftwopt}) has short distance divergences which arise when the two points approach each other, ${\bf x} \rightarrow {\bf y}$. These need 
 to be regulated in order to make the calculation well defined. One way to do so is to work  with the probability  not for electric fields at points $E_r({\bf x})$ but instead for electric fields which are smoothed out over a small  distance  
 scale. Here, instead we will take the two points ${\bf x}, {\bf y}$ to lie on two different spheres of radius $R, R'$ respectively, with, $R'=R+ \Delta$, where 
 \begin{equation}
 \label{valdelta}
  \Delta \ll R.
  \end{equation}
  This turns out to be equivalent, for our purposes, to the smoothing out procedure and is  easier to implement.
  
  We now have two small distance cut-offs that have been introduced, $\Delta$ above and $\epsilon$ which appears in the 
  entanglement entropy eq.(\ref{see}). 
  The spherical symmetry of this problem ensures that the operator $G_{rr}^{-1}$ is diagonal in the spherical harmonic basis. The short distance cut-off leads to a maximum value for the angular momentum, $l_{max}$, of the modes that are being included.  
  We take 
  \begin{equation}
  \label{deflm}
  l_{max}\sim {R\over \epsilon},
  \end{equation}
  so that in effect $\Delta$ is the cut-off along the radial direction, whereas $\epsilon$ is the cut-off in the angular directions
  on the sphere. Of course,  if the underlying regulator is a lattice of the kind we have considered above,  the two cut-offs would  be the same, but it is convenient for our case, having introduced them as distinct, to instead consider the limit, 
  \begin{equation}
  \label{limde}
  \Delta \ll\epsilon.
  \end{equation}
   This amounts to keeping the effects of  modes $l< l_{max}$ where
   \begin{equation}
   \label{relmax}
   l_{max} \ll {R\over \Delta}.
   \end{equation}
   We will see that imposing the condition eq.(\ref{relmax}) on the modes will simplify the calculations. 
   The $\log(A)$ term gets contributions from a range of $l$ and its coefficient can be reliably obtained  by looking at the range which meets eq.(\ref{relmax}).

   The Greens function, eq.(\ref{deftwopt}), can be calculated, as described in Appendix \ref{appendix:greens-fn}. We get, 
  \begin{equation}
 \label{twopt}  G_{rr} = - \frac{1}{\pi^{2} (R^{2} + R'^{2})^{2}} \frac{\alpha - \cos \gamma}{(1-\alpha \cos \gamma)^{3}},
 \ee 
 where $\gamma$ is the angle between the two points
 \begin{equation}
 \label{defgamma}
 \cos \gamma= {\hat x} \cdot {\hat y}
 \end{equation}
 and 
 \begin{equation}
 \label{defalpha} 
  \alpha = \frac{2RR'}{R^{2} + R'^{2}} = 1 - \frac{\Delta^{2}}{2 R^2}.
\end{equation}

To express this in the spherical harmonic basis we expand $G_{rr}$ in a power series in $\cos\gamma$ and then use the relation
\begin{equation}
\label{spha} 
 P_{l} (\cos\gamma) = \frac{4\pi}{2l+1} \sum_{m} Y_{l}^{m} (\theta_{1},\phi_{1}) Y_{l}^{m*} (\theta_{2},\phi_{2}).\end{equation}
Note that the power series expansion  in $\cos\gamma$ is valid when 
\begin{equation}
\label{pws}
\alpha<1,
\end{equation}
from eq.(\ref{defalpha}) this requires, 
\begin{equation}
\label{red}
\Delta\ne 0.
\end{equation}
It is also clear that the resulting power series will have a divergence when $\alpha\rightarrow 1$. 
We will be interested in small values of $\Delta/R$ and thus in the leading divergence which arises in this limit. 

Some of the resulting algebra is described in more detail in appendix \ref{appendix:greens-fn}.
Working  self-consistently in the limit where eq.(\ref{relmax})  is met we get that the leading divergence is logarithmic 
going like $\log({R^2 \over \Delta^2})$, and gives rise to a contribution
\begin{equation}
\label{cong}
G_{rr}^{lm} = \frac{1}{\pi R^{4}}  \left( \log \frac{R^{2}}{\Delta^{2}} \right) l(l+1) .
  \end{equation}
 Additional terms in $G_{rr}$ are sub-dominant when   $\Delta/R\ll 1$.

 Notice that the Green's function is proportional simply to the  two-dimensional Laplacian on $S^2$.  Using standard heat kernel methods we then get that 
 \begin{equation}
 \label{contg}
 {1\over 2} \log \det G_{rr}^{-1} = - {1\over 3} \log({R\over \epsilon})+ \cdots
 \end{equation}
The coefficient, ${1\over 3}$,  is determined by the central charge of the two dimensional free  scalar field theory. 
 The ellipses in eq.(\ref{contg}) denote additional terms   which do not contribute to the log term. The prefactor, 
 $\frac{1}{\pi R^{4}}  \log \frac{R^{2}}{\Delta^{2}}$,  gives rise to  a term going like $A/\epsilon^2$ in eq.(\ref{contg}).
  
 Neglecting the $D_4$ term in eq.(\ref{constn})we then see from eq.(\ref{ret}) that 
 \begin{equation}
 \label{retb}
-\sum_ip_i\log(p_i)= - {1\over 6} \log({A\over \epsilon^2}) + \cdots
\end{equation}
where the ellipses denote terms  which do not contribute to the $\log$ term of interest, and $A$ and $R$ are related by eq.(\ref{vala}). 

Putting all this together we finally get that the logarithmic contribution to the extractable part of the entanglement goes like,
\begin{align}
S_{extract} & = [ -{31\over 90}+ {1\over 6}]\log({A \over \epsilon^2})\label{extracs} \\
&= -{16\over 90}\log({A \over \epsilon^2}), \label{extracb}
\end{align}
which agrees with eq.(\ref{valD}). 
As  mentioned above, that this is different from the full entanglement entropy.
\subsection{Comments}
Let us end this section  with some comments. 
We begin by addressing some of the points in the calculation above more carefully. 

We have not been careful about the exact definition of the normalisation,  $N$,  eq.(\ref{pe}), which in turn in tied to the measure for the functional integral. Ambiguities in defining this measure can be absorbed into different choices of local counter terms on the two-dimensional boundary.  These can change the coefficient of the leading area law divergence and also the finite terms but not the coefficient of the log term. One can think of these as changes in the coefficient $D_4$ or the non-log pieces in $\log\det G_{rr}^{-1}$ of eq.(\ref{constn}). This is also clear  from our final result which is expressed in terms of the determinant of the two-dimensional  Laplacian. This determinant has ambiguities related to the two-dimensional cosmological constant, etc. These give rise to a change in the coefficient of the area term but not the coefficient of the log piece. 
 
 A precise definition for $N$ will arise in any  well defined way to regulate the theory, e.g. if we start with the lattice definition used above. It is worth pointing out that the normalisation constant in this case will depend on the coupling constant $g$. The link variables $\theta_{ij}$ on the lattice are compact, $\theta_{ij}\in [0,2\pi]$, as a result their conjugate variables
$L_{ij}$, are  quantised with integer eigenvalues, which we denote as $n$ here. 
From the lattice and continuum actions, eq.(\ref{actu1}), eq.(\ref{maxwell}) and the Hamiltonian, eq.(\ref{defH}) it follows,  that 
\begin{equation}
\label{relEL}
E_j({\bf x}_i)= {g \over \epsilon^2} {\cal L}_{ij}.
\end{equation}
where $E_j({\bf x}_i)$ is the electric field along the ${\hat j}$ direction emanating from\footnote{$\epsilon$ in eq.(\ref{relEL})
is a lattice cutoff, we are not being careful here about the cutoffs in the spatial and temporal directions which can be different.} ${\bf x}_i$. 
As a result the sum, 
\begin{equation}
\label{vals}
\sum_n \rightarrow \int dE_j {\epsilon^2 \over g}
\end{equation}
and this gives rise to the measure for the sum over the electric fields,  
\begin{equation}
\label{mesele}
\int D[E_r] = \prod_{  { \bf x_i},j} dE_j({\bf x}_i){\epsilon^2 \over g }
\end{equation}
where to define the product on the RHS we are considering discrete values of ${\bf x}_j$ valued on a spatial lattice of size $\epsilon$. Here the spatial lattice lies on the $S^2$ boundary. 
For small enough $g$ we see from eq.(\ref{vals})  that it is a good approximation to replace the sum over integer $n$ by a continuous integral
(analogous to the sum over discrete momenta being replaced by $\int {dp \over 2 \pi \hbar}$ for a free particle). 
We also see  from eq.(\ref{mesele}) that the normalisation $N$ must then go like $N \sim g^{N_b}$, since $N_b\sim A/\epsilon^2$ is the total number of points on the $S^2$ boundary. 
From eq. (\ref{ret})  we now see that this dependence on $g$ givers rise to a contribution
\begin{equation}
\label{contaa}
\Delta[-\sum_i p_i \log p_i ]\sim -\log (g) {A \over \epsilon^2}.
\end{equation}

More precisely, since the Gauss law constraint must be met and $\int E_r d\Omega=0$, there are $N_b-1$ number of independent normal components of the boundary electric field. Thus $N\sim g^{N_b-1}$ leading to 
\begin{equation}
\label{comtat}
 \Delta[-\sum_i p_i \log p_i ]=-(N_{b}-1) \log(g)
 \end{equation}
 The reduction by  $-1$ in the prefactor, is analogous to what happens in the toric code model for $2+1$ dimensional discrete gauge theories \cite{Kitaev2005,Levin2005}  as has been emphasised in \cite{Pretko2015}.
 
 As a result of this reduction, the topological entanglement proposed in \cite{Grover2011}, which is a generalisation of the topological entanglement in $2+1$ dimensions, proposed in \cite{Kitaev2005,Levin2005}, acquires a contribution, 
\begin{equation}
 \label{stopc}
 \Delta S_{top}=-\log(g).
 \end{equation}
 Note however that eq.(\ref{stopc}) is not the full result for $S_{top}$ in the $U(1)$ theory \footnote{Also note that in eq.(\ref{stopc}) $g$ is the coupling constant, unlike \cite{Pretko2015} who obtain a contribution going like $\log(L)$, where $L$ is the size of the region.}.  There are additional contributions,
  since there are massless excitations in the system,  and these  contributions   are in fact non-topological, changing  under smooth deformation of the three regions involved in the  definition of the topological entanglement. 
 This is clear, for example, from the classical term above, eq.(\ref{cong}), which depends on the two dimensional  massless scalar Laplacian on the boundary.

In fact the Gauss law constraint is also important for understanding the zero modes, to which we  turn next.   
From our result, eq.(\ref{cong}), we see that the contribution of $l=0$ mode to the two point function vanishes. 
This also follows from Gauss' law since  the integral $\int E_r d\Omega=0$ on $S^2$ must vanish, a fact which  can be directly verified from eq.(\ref{twopt}) as well.\footnote{It is important for this check to work, that $\Delta \ne 0$ and the two points have been separated in the radial direction.}
As a result, more precisely, the determinant in eq.(\ref{contg}) has been  evaluated over the non-zero modes, $l\ne 0$. 

The fact that the  two dimensional scalar  Laplacian appears in eq.(\ref{cong}) is a striking fact and can be argued  to be true more generally as well. Consider any region $R$, whose entanglement is of interest. Then one can argue 
the leading contribution to the Green's function for the  normal component of ${\bf E}$ will arise from the scalar  Laplacian on the boundary of $R$. 
To see this let us first redo the calculations above in a somewhat different way which makes the appearance of the two-dimensional Laplacian for the $S^2$ case more transparent. 

As discussed in appendix \ref{appendix:greens-fn}, the two point function, eq.(\ref{twopt}) , can be written as 
\begin{align}
G_{rr}& ={2\over \pi} \left[  {1\over 12} { (R^2-R'^2)^2 \over 2 R R' } \sum_{l,m}\int dk k^5 j_l(kR) j_i(kR') Y_{lm}(\theta,\phi)
Y^*_{lm}(\theta',\phi') \right. \label{twoptaa}\\
& \quad \quad \quad \quad \quad + \left. {1\over \alpha} \sum_{lm} \int dk k^3 j_l(kR) j_l(kR') Y_{lm}(\theta,\phi) Y^*_{lm}(\theta',\phi')\right]\nonumber
\end{align}
where $\alpha$ is given in eq.(\ref{defalpha}). 

Note that $\phi_{klm}=j_l(kr) Y_{lm} (\theta, \phi)$ is an eigenvalue  of  the scalar Laplacian in $3$ dimensions,
\begin{align}
\label{sols}
\nabla^2 \phi_{klm} &=  -k^2 \phi_{klm}\\
{1\over r^2}\partial_r (r^2 \partial_r \phi_{klm})-{l(l+1)\over r^2} \phi_{klm} &= -k^2 \phi_{klm}
\end{align}
with $l(l+1)$ being the eigenvalue of the two dimensional Laplacian on $S^2$. 

In the limit, eq.(\ref{relmax}), the contribution to the sum, eq.(\ref{twoptaa}), is dominated by modes with  radial momentum bigger than the momentum along the $S^2$  boundary, $k\gg {l\over R}$.

We can use the WKB approximation to understand the behaviour of the modes in this limit. In this approximation  a solution goes like,
\begin{equation}
\label{lwkb}
\phi \sim e^{i \pm \int dr \sqrt{E-V}},
\end{equation}
 where $E=k^2r^2$ and the potential term arises from the two dim. Laplacian,  $V=l(l+1)/r^2$. The leading term in eq.(\ref{lwkb}) comes from neglecting $V$ and goes like $e^{\pm i (kr + \theta)}$, 
where $\theta$ is a phase. The next term comes from expanding the square root $\sqrt{E-V}\simeq \sqrt{E}[1-{1\over 2}
{V \over E}]$, and is proportion to $l(l+1)$, the eigenvalue of the two-dim Laplacian. 

Done more carefully, this  gives rise to the standard asymptotic expansion \cite{ArfkenBook},
\begin{align}
\label{behj}
j_l(kr)&=\sqrt{\pi \over 2 kr}[H^{(1)}_{l+1/2}(kr)  + c.c.] \nonumber \\
& \sim \{ e^{i kr - \theta}[P_{l+{1\over 2}}(kr) +i Q_{l+{1\over 2}}(kr) ] + c.c.\}.
\end{align}

The leading behaviour which comes from setting $Q_{l+{1\over 2}}=0, P_{l+{1\over 2}}=1$
cancels out in the two integrals in eq.(\ref{twoptaa}). 
The first non-trivial contribution then comes from keeping the sub-leading term.
From arguments given above it follows that its coefficient is therefore proportional to the eigenvalue of the two dim. Laplacian. In addition, 
it is logarithmically divergent going like, 
\begin{equation}
\label{ldiv}
\int {dk \over k} e^{ik(R-R')} \sim \log({R \over \Delta }).
\end{equation}
We see that this reproduces the result, eq.(\ref{cong}) up to an overall constant, which we did not keep track of.

These arguments  make it clear that in the more general case as well, since  we are  working in the limit where the  component of the momentum normal to the boundary is much bigger than the component along the boundary, the final result for the 
greens function, eq.(\ref{cong}), will be given by the two dimensional Laplacian on the boundary, $\nabla_B^{2} $ multiplied by a logarithmic divergence, 
\begin{equation}
\label{congc}
G_{rr}\propto \log({R^2\over \Delta^2}) (-\nabla_B^{2}).
\end{equation}
Thus $\det(\nabla^{2}_B)$ will determine the difference between the full and extractable entanglement entropy, eq.(\ref{cong}), eq.(\ref{contg}).

Finally, let us note that the results above help us understand some of the discrepancies in the literature. 
Both results, eq.(\ref{finvalc})  and eq.(\ref{extracb}) have been obtained earlier, for the coefficient of the log term, when considering the $U(1)$ theory. We see that the first result, eq.(\ref{finvalc}),which follows from the $a$ anomaly coefficient is the total entanglement in the extended Hilbert space definition and also the Replica trick that follows from it.  The second, eq.(\ref{extracb}), is the extractable part  corresponding to the  number of Bell pairs which can be distilled from the system, etc. 
The second result is tied to physical measurements and thus independent of definitions. 
 We also note that  naively speaking one might think that the $U(1)$ theory would give the same result as two massless scalars in $3+1$ dimensions, since there are two transverse modes for a photon. However using the $a$ anomaly coefficient for the massless scalar 
 we get that two scalars would give a log term,
 \begin{equation}
 \label{twos}
 S_{EE}=2\times (-{1\over 180}) \log({A \over \epsilon^2})=-{1\over 90}\log({A\over \epsilon^2}).
 \end{equation}
 This does not agree with either  of the two results above,  for $S_{EE}$, eq.(\ref{finvalc}),  or $S_{ext}$, eq.(\ref{extracb}).
 The fact that the  gauge theory result  could differ from that obtained from two  scalars was first noted in \cite{Kabat1995} and is related to the presence of extra terms in the path interval, called ``Kabat terms''.

 \section{Conclusions} \label{sec:conclusion}
 In this paper we have explored some features of the extended Hilbert space definition of entanglement entropy in gauge theories by focusing on a simple example of the free $U(1)$ theory in $3+1$ dimensions in the continuum.
 It has been noted earlier that this definition agrees with the replica trick method of calculating the entanglement \cite{Ghosh2015,Aoki2015}. 
 It has  also been noted \cite{Soni2015,vanAcoleyen2015}  that the extended Hilbert space definition differs from the  extractable entanglement, which is the maximum number of Bell pairs that can be obtained in entanglement distillation  or available for entanglement dilution, and the difference between the two was  precisely stated in \cite{Soni2015,vanAcoleyen2015}. 
 
 Here we start with the $U(1)$ theory on the lattice at weak coupling and take the  limit carefully to arrive at the continuum limit of the path integral needed for the replica trick. The path integral is  gauge-invariant with the gauge-fixing delta function being accompanied with the required Faddeev-Popov determinant. 
 We then calculate both the full entanglement, as given in the extended Hilbert space definition, and the extractable piece. 
 More precisely we calculate the coefficient of the log term, eq.(\ref{see}), in these cases. We find that the two are different. 
 While the full entanglement has a coefficient $C=-{31 \over 90}$ the extractable piece has coefficient $D=-{16\over 90}$, eq.(\ref{finvalc}) and eq.(\ref{extracb}), eq.(\ref{valD}).
  The difference is related to the central charge of a masses scalar that lives on the two-dimensional boundary. 
  We also argued that this is a general feature. For any region ${\cal R}$ in this theory the two kinds of entanglement will differ and the difference will be related to the determinant of the Laplacian for a massless scalar living on the two dimensional boundary of ${\cal R}$. 
  
  We hope our analysis has helped resolve some of the differences in the literature where both results, $C$, eq.(\ref{finvalc}), and $D$, eq.(\ref{valD}), for the coefficient of the log term in the entanglement have been obtained.  We see that these difference arise because the two calculations pertain to two different quantities. 
  
  Let us also comment that on the lattice we identify a contribution to the entanglement entropy, which arises due to the Gauss law constraint. This contribution  given in eq. \eqref{stopc}, goes like $\log(g)$, where $g$ is the coupling constant of the theory,  and is the analogue of a term known to arise in the toric code model for discrete gauge theories \cite{Pretko2015}. In turn, this contribution to the entanglement entropy gives rise to a term  in the topological entanglement entropy, $S_{top}$, eq. \eqref{stopc},  although there are other contribution  to $S_{top}$ as well, due to the presence of massless degrees of freedom.
  
  It is also worth mentioning that the $U(1)$ gauge symmetry   could arise  as the low-energy limit of a more complete theory. 
  In this case it could be that the full entanglement is extractable using operators or excitations present in the full theory but not in its low-energy limit. This could happen, for example, in condensed matter systems like quantum spin  liquids  some of which are known to give rise to a free $U(1)$ gauge theory in the infrared. Even in such cases the difference found above between the full entanglement and its extractable piece is interesting, since the extractable piece tells us about the Bell pairs which can be obtained using only low-energy probes which couple to gauge invariant operators. 
    
    The entanglement entropy in the extended Hilbert space definition agrees with the electric centre choice \cite{Casini2013,Ghosh2015}. 
  Other definitions for the entanglement can also be given, these will not agree with the replica trick path integral we have obtained. 
  On the other hand, the extractable entanglement is physical , since it has an operational significance in terms of extractable Bell pairs. 
  We also expect, for the same reasons, that the extractable entanglement is invariant under electro-magnetic duality.

  It is also worth commenting, as emphasised in \cite{Casini2013,Casini2014}, that other quantities like the mutual information or the relative entropy are less sensitive to the choice of the centre than the entanglement entropy itself. For  the   $U(1)$ theory considered here it was shown in \cite{Casini2014} by a numerical analysis that the dependence on  the choice of the centre   drops out for  the mutual information in the continuum limit.

  One direction is which these results should be generalised is to consider non-Abelian theories. 
  The difference between the two kinds of entanglement in this case has an extra term, tied to the fact that irreducible representations in the non-Abelian case have dimensions greater than unity. One expects that the difference between the two kinds of entanglement can be expressed in terms of a contribution arising from the  boundary  of the region of interest in this case as well. 
  
  It is also important to connect this discussion to gravity. The Ryu-Takayanagi entanglement \cite{Ryu2006} in AdS gravity, 
  which corresponds to a minimal area surface,  has been shown to follow from the replica trick in the boundary \cite{Lewkowycz2013}. Since we have argued   that the replica trick is equivalent to the extended Hilbert space definition, it follows in turn that the Ryu-Takayanagi entanglement in the bulk agrees with this definition.
 However, we have seen that the extractable entanglement, which has a clear physical significance, is different in general. 
 It will be very interesting to ask what the difference corresponds to on the gravity side and whether it can be expressed in terms of geometric quantities, for example related to minimal area surfaces or degrees of freedom living on such surfaces.


 \section{Acknowledgments}
 We thank Y. Dandekar, N. Iizuka, G. Mandal, S. Minwalla, P. Nayak, \DJ.~Radi\v cevi\'c, A. Saha, A. Sen, T. Senthil, D. Tong, T. Takayanagi for discussions. 
 SPT acknowledges support from the J. C. Bose fellowship of the Government of India. 
 We acknowledge support from the Department of Atomic Energy, Government of India.
 Most of all we thank the people of India for generously supporting research in string theory. 
  
\appendix 
\section{Calculation of the Green's Function} \label{appendix:greens-fn}
In this section, we will calculate the two-point function $G_{rr}$ of the radial component of the electric field on the sphere, and from there prove that its contribution to the log term in the entropy is $-{1\over 6} \log A/\varepsilon^{2}$, eq. \eqref{retb}.

The strategy for finding the Green's function will be to find the two-point function in momentum space, using the standard quantisation rules, then Fourier transform it to position space, and then finally decompose this answer in terms of spherical harmonics.
Then, $\log \det G_{rr}$ can be calculated using a standard heat kernel expansion. 

The vector potential is quantised, in $A_{0} = 0$ and $\vec{\grad} \cdot \vec{A} = 0$ gauge, as
\begin{equation}
  A_{i} = \int \frac{d^{3}k}{(2\pi)^{3}} \frac{1}{\sqrt{2k}} \sum_{\alpha} \left\{ a_{\bk,\alpha} \epsilon^{\alpha}_{i} (\bk) e^{-i (kt - \bk \cdot \bx)} + a_{\bk,\alpha}^{\dagger} \epsilon^{\alpha *}_{i} (\bk) e^{i (kt - \bk \cdot \bx)} \right\}, \quad \sum_{\alpha} \epsilon^{\alpha}_{i} (\bk) \epsilon^{\alpha *}_{i} (\bk) = \delta_{ij} - \frac{k_{i} k_{j}}{k^{2}}.
  \label{eqn:A-expansion}
\end{equation}
Because of this, the electric field is 
\begin{equation}
  E_{i} = -i \int \frac{d^{3} k}{(2\pi)^{3}} \sqrt{\frac{k}{2}} \sum_{\alpha} \left\{ a_{\bk,\alpha} \epsilon^{\alpha}_{i} (\bk) e^{-i (kt - \bk \cdot \bx)} - a_{\bk,\alpha}^{\dagger} \epsilon^{\alpha *}_{i} (\bk) e^{i (kt - \bk \cdot \bx)} \right\}
  \label{eqn:E-expansion}
\end{equation}
and the Green's function is
\begin{equation}
  \langle E_{i} (\bx) E_{j} (\by) \rangle = \frac{1}{2} \int \frac{d^{3}k}{(2\pi)^{3}} k \left( \delta_{ij} - \frac{k_{i} k_{j}}{k^{2}} \right) e^{i \bk \cdot (\bx-\by)}.
  \label{eqn:greens-fn-mom-space}
\end{equation}

The classical contribution is given by the two-point function with both points on the sphere.
However, naively choosing both points on the sphere gives un-physical divergences, including the monopole term not vanishing as it should because of Gauss' law.
So, we regularise by smoothing out the electric field a little in the radial direction; we choose the two points to be on spheres of radii $R_{1} = R$ and $R_{2} = R + \Delta$ respectively.
Since we want the two spheres to be coincident, we take the spacing between the spheres to be much smaller than the lattice scale, $\Delta \ll \varepsilon$.

We can now proceed to calculate this regularised Green's function.
Defining $\xi = \bx - \by$, we rewrite the Green's function as
\begin{align}
  G_{rr} (\bxi) &= \frac{1}{2 (2\pi)^{2}} \int dk d(\cos\theta) \left( k^{3} \hat{x} \cdot \hat{y} - k \bk \cdot \hat{x} \bk \cdot \hat{y} \right) e^{ik\xi\cos\theta}.
  \label{eqn:greens-fn-again}
\end{align}
The first term is
\begin{align}
  \frac{\hat{x} \cdot \hat{y}}{8\pi^{2}} \int_{0}^{\infty} dk k^{3} \int_{-1}^{1} d(\cos\theta) e^{ik\xi\cos\theta} &= \frac{\hat{x} \cdot \hat{y}}{8\pi^{2}} \int dk k^{3} \frac{e^{ik\xi} - e^{-ik\xi}}{ik\xi} \nonumber\\
  &= \frac{\hat{x} \cdot \hat{y}}{4\pi^{2}\xi} \left( - \partial_{\xi}^{2} \right) \Im \int dk e^{ik\xi} \nonumber\\
  &= - \frac{\hat{x} \cdot \hat{y}}{2\pi^{2}\xi^{4}}.
  \label{eqn:greens-fn-first-term}
\end{align}
And the second term is
\begin{align}
  - \frac{1}{8\pi^{2}} \int dk d(\cos\theta) k \bk \cdot \hat{x} \bk \cdot \hat{y} e^{i\bk\cdot\bxi} &= - \frac{1}{8\pi^{2}} \left( - \hat{x} \cdot \vec{\grad}_{\xi} \hat{y} \cdot \vec{\grad}_{\xi} \right) \int dk k \int d(\cos\theta) e^{ik\xi\cos\theta} \nonumber\\
  &= - \frac{1}{8\pi^{2}} \left( - \hat{x} \cdot \vec{\grad}_{\xi} \hat{y} \cdot \vec{\grad}_{\xi} \right) \frac{2}{\xi} \Im \int dk e^{ik\xi} \nonumber\\
  &= \frac{\hat{x}_{i} \hat{y}_{j}}{4\pi^{2}} \partial_{i} \partial_{j} \frac{1}{\xi^{2}} \nonumber\\
  &= - \frac{1}{2\pi^{2}\xi^{4}} \left( \hat{x} \cdot \hat{y} - 4 \frac{\bxi \cdot \hat{x} \bxi \cdot \hat{y}}{\xi^{2}} \right).
  \label{eqn:greens-fn-second-term}
\end{align}
Calling the angle between the points $\gamma$, the various inner products above are
\begin{align}
  \hat{x} \cdot \hat{y} &= \cos \gamma \nonumber\\
  \bxi \cdot \hat{x} &= R_{1} - R_{2} \cos \gamma \nonumber\\
  \bxi \cdot \hat{y} &= R_{1} \cos \gamma - R_{2} \nonumber\\
  \xi^{2} &= R_{1}^{2} + R_{2}^{2} - 2R_{1}R_{2} \cos \gamma.
  \label{eqn:ip-simplifications}
\end{align}
So, the Green's function is
\begin{equation}
  G_{rr} = - \frac{1}{\pi^{2} (R_{1}^{2} + R_{2}^{2})^{2}} \frac{\alpha - \cos \gamma}{(1-\alpha \cos \gamma)^{3}}, \quad \alpha = \frac{2R_{1}R_{2}}{R_{1}^{2} + R_{2}^{2}} = 1 - \frac{1}{2} \frac{\Delta^{2}}{R^{2}}.
  \label{eqn:greens-fn-real-space}
\end{equation}

To diagonalise it, we will expand this in a basis of Legendre functions and use the relation
\begin{equation}
  P_{l} (\cos\gamma) = \frac{4\pi}{2l+1} \sum_{m} Y_{l}^{m} (\theta_{1},\phi_{1}) Y_{l}^{m*} (\theta_{2},\phi_{2}).
  \label{eqn:pl-to-ylm-relation}
\end{equation}

To expand it in terms of spherical harmonics, we first expand out the denominator to write it as a power series in $\alpha$,
\begin{align}
  G_{rr} &= - \frac{1}{\pi^{2} (R_{1}^{2} + R_{2}^{2})^{2}} (\alpha-\cos\gamma) \sum_{n=0}^{\infty} \frac{(n+1)(n+2)}{2} \alpha^{n} \cos^{n} \theta \nonumber\\
  &= \frac{1}{\pi^{2} (R_{1}^{2} + R_{2}^{2})^{2}} \sum_{n=0}^{\infty} \frac{n+1}{2} \left[ \frac{1-\alpha^{2}}{\alpha} n - 2\alpha \right] \alpha^{n} \cos^{n} \gamma,
  \label{eqn:grr-expansion-in-powers}
\end{align}
and then use the relations \cite{GradshteynBook}
\begin{align}
  t^{2n} &= \sum_{k=0}^{\infty} (4k+1) \frac{2n!!}{(2n-2k)!!} \frac{(2n-1)!!}{(2n+2k+1)!!} P_{2k} (t) \nonumber\\
  t^{2n+1} &= \sum_{k=0}^{\infty} (4k+3) \frac{2n!!}{(2n-2k)!!} \frac{(2n+1)!!}{(2n+2k+3)!!} P_{2k+1} (t).
  \label{eqn:powers-as-pls}
\end{align}

We plug eqn \eqref{eqn:powers-as-pls} into eqn \eqref{eqn:grr-expansion-in-powers},
\begin{align}
  \pi^{2} (R_{1}^{2} + R_{2}^{2})^{2} G_{rr} 
  &= \sum_{k=0}^{\infty} (4k+1) \left\{ \sum_{n=0}^{\infty} \left( \frac{1-\alpha^{2}}{\alpha} n - \alpha \right) \frac{2n!!}{(2n-2k)!!} \frac{(2n+1)!!}{(2n+2k+1)!!} \alpha^{2n} \right\} P_{2k} \nonumber\\
  &\quad + \sum_{k=0}^{\infty} (4k+3) \left\{ \sum_{n=0}^{\infty} \left( \frac{1-\alpha^{2}}{\alpha} n - \frac{3\alpha^{2}-1}{2\alpha} \right) \frac{(2n+2)!!}{(2n-2k)!!} \frac{(2n+1)!!}{(2n+2k+3)!!} \alpha^{2n+1} \right\} P_{2k+1}.
  \label{eqn:grr-in-pls}
\end{align}
Note that the factors of $4k+1$ and $4k+3$ will exactly cancel those that come from converting the Legendre polynomial into spherical harmonics.

The part of the sum contributing to its divergence is
\begin{equation}
  n \sim \frac{1}{1-\alpha^{2}} = \frac{R^{2}}{\Delta^{2}}.
  \label{eqn:divergent-part-of-sum}
\end{equation}
Since the maximum angular momentum allowed is
\begin{equation}
  l_{max} \sim \frac{R}{\varepsilon} \ll \frac{R}{\Delta} \ll \frac{R^{2}}{\Delta^{2}},
  \label{eqn:angular-momentum-cutoff}
\end{equation}
as long as we're interested in only the divergent pieces, we can safely work in the regime
\begin{equation}
  k \ll n
  \label{eqn:sum-divergence-regime}
\end{equation}
and look at terms order by order in a $1/n$ expansion.

To do this expansion, we rewrite the double factorials as
\begin{align}
  \frac{2n!!}{(2n-2k)!!} \frac{(2n+1)!!}{(2n+2k+1)!!} &= \prod_{r=0}^{k-1} \left( 1 - \frac{1}{2} \frac{2k+1}{n+k+ \frac{1}{2} - r} \right) \quad \text{and} \label{eqn:even-dbl-factorial}\\
  \frac{(2n+2)!!}{(2n-2k)!!} \frac{(2n+1)!!}{(2n+2k+3)!!} &= \prod_{r=-1}^{k-1}  \left( 1 - \frac{1}{2} \frac{2k+1}{n+k+ \frac{1}{2} - r} \right).\label{eqn:odd-dbl-factorial}
\end{align}

Clearly, in both terms the RHS goes as $1 + 1/n + 1/n^{2} + \cdots$ at large $n$.
Now, there are two terms multiplying the double factorial, in eq.(\ref{eqn:grr-in-pls}) 
one proportional to $n$ and one proportional to $1$.
Thus, the even term in eq.(\ref{eqn:grr-in-pls}) splits into two terms, one of which goes as $n + 1 + 1/n + \cdots$ and the other of which goes as $1 + 1/n + 1/n^{2} + \cdots$, and similarly for the odd term.
Naively, then, the leading divergence is quadratic.
However, the term proportional to $n$ comes with a factor of $1-\alpha^{2}$, which reduces the power of the divergence by one order, and so the leading divergence is linear.

In fact, the linear divergence actually cancels.
The terms corresponding to the leading divergences all come from when the term which is only a product of $1$s for every $r$ in eqns. \eqref{eqn:even-dbl-factorial} or \eqref{eqn:odd-dbl-factorial}.
For the even term, this part is
\begin{equation}
  \frac{1-\alpha^{2}}{\alpha} \sum n\alpha^{2n} - \alpha \sum \alpha^{2n} = \frac{1-\alpha^{2}}{\alpha} \frac{\alpha^{2}}{(1-\alpha^{2})^{2}} - \alpha \frac{1}{1-\alpha^{2}} = 0.
  \label{eqn:even-lin-div}
\end{equation}
Similarly, for the odd term, this part is
\begin{equation}
  \frac{1-\alpha^{2}}{\alpha} \sum n \alpha^{2n+1} - \frac{3\alpha^{2} - 1}{2\alpha} \sum \alpha^{2n+1} = \frac{1-\alpha^{2}}{\alpha} \frac{\alpha^{3}}{(1-\alpha^{2})^{2}} - \frac{3\alpha^{2} - 1}{2\alpha} \frac{\alpha}{1-\alpha^{2}} = \frac{1}{2},
  \label{eqn:odd-lin-div}
\end{equation}
which, while not $0$ per se, is regular.

So, the leading divergence is logarithmic.
Only the part that didn't have an $n$ multiplying the double factorials in eq.(\ref{eqn:grr-in-pls})  can contribute to the log divergence, since the log divergent piece in the part with the $n$ vanishes because of the multiplication with $1-\alpha^{2}$.
In the part without the $n$, the first sub-leading term is the one where exactly one of the factors in the products \eqref{eqn:even-dbl-factorial} or \eqref{eqn:odd-dbl-factorial} doesn't contribute a $1$, resulting in a piece that is overall of $O(1/n)$.
For the even term, it is
\begin{align}
  \alpha \sum_{n} \sum_{r = 0}^{k-1} \frac{2k+1}{2n+2k+1-2r} \alpha^{2n} &= \frac{\alpha}{2} (2k+1) \sum_{r=0}^{k-1} \sum_{n} \left( \frac{\alpha^{2n}}{n} + O (1/n^{2}) \right) \nonumber\\
  &= \frac{l(l+1)}{4} \log \left( \frac{1}{1-\alpha^{2}} \right).
  \label{eqn:even-log-div}
\end{align}
In obtaining the second line  on the RHS we have set $2k=l$, and also set the prefactor $\alpha$ outside the sum to be unity. 
And for the odd term it is
\begin{align}
  \frac{3\alpha^{2}-1}{2\alpha} \sum_{n} \sum_{r=-1}^{k-1} \frac{2k+1}{2n+2k+1-2r} \alpha^{2n+1} &= \frac{3\alpha^{2}-1}{2\alpha} {2k+1\over 2}\sum_{r=-1}^{k-1} \sum_{n} \left( \frac{\alpha^{2n+1}}{n} + O(1/n^{2}) \right) \nonumber\\
  &= \frac{l(l+1)}{4} \log \left( \frac{1}{1-\alpha^{2}} \right).
  \label{eqn:odd-log-div}
\end{align}
Once again, in obtaining the second line on the RHS we set $\alpha=1$ in the  prefactor multiplying the sum and also set 
$2k+1=l$. 

All the rest of the terms do not  contribute to the log part and we ignore them.
Thus, putting $R_{1} = R$ and $R_{2} = R+\Delta$, the Green's function to leading order is
\begin{equation}
  G_{rr}^{lm} = \frac{1}{\pi R^{4}} l(l+1) \log \frac{R^{2}}{\Delta^{2}}.
  \label{eqn:greens-fn-final}
\end{equation}

As noted in the main text, this means that the Green's function to leading order is $\grad^{2}$, where $\grad^{2}$ is the Laplacian on the two-dimensional sphere.
Thus, the entropy is $\frac{1}{2} \log \det \grad^{2}$.
To evaluate this, we use a heat kernel expansion
\begin{equation}
  \frac{1}{2} \tr \log \grad^{2} = - \frac{1}{2} \int_{\varepsilon^{2}}^{\infty} \frac{dt}{t} \tr e^{t \grad^{2}}.
  \label{eqn:heat-kernel-expansion}
\end{equation}
For two-dimensional manifolds without boundary, the short-time asymptotic expansion of the heat kernel for the Laplacian is known to be (see for example \cite{ParkerBook})
\begin{equation}
  \tr e^{t\grad^{2}} \approx \frac{1}{4\pi t} \left\{ \tr \mathds{1} + t \tr \left( \frac{R}{6} \mathds{1} \right) \right\},
  \label{eqn:heat-kernel-short-t-exp}
\end{equation}
where $R$ is not the radius but the Ricci scalar.

Clearly, it is the second term that gives a log divergence.
Substituting the Ricci scalar in terms of the radius as $2/R^{2}$ and $\tr \mathds{1} = 4 \pi R^{2}$, we get the log divergent piece to be
\begin{equation}
  - \frac{1}{6} \log \frac{R^{2}}{\varepsilon^{2}}.
  \label{eqn:log-term}
\end{equation}
This agrees with eq.(\ref{contg}) above. 


\begin{thebibliography}{99}
\bibitem{Ghosh2015}
S.~Ghosh, R.~M Soni, S.~P. Trivedi, ``On the Entanglement Entropy for Gauge Theories,''
\href{http://dx.doi,org/10.1007/JHEP09(2015)069}{\emph{JHEP} \textbf{1509} (2015) 069},
\href{http://arxiv.org/abs/1501.02593}{{\ttfamily arXiv:1501.02593 [hep-th]}}.

\bibitem{Aoki2015}
S.~Aoki, T.~Iritani, M.~Nozaki, T.~Numasawa, N.~Shiba, H.~Tasaki, ``On the Definition of Entanglement Entropy in Lattice Gauge Theories,''
\href{http://dx.doi.org/10.1007/JHEP06(2015)187}{\emph{JHEP} \textbf{1506} (2015) 187},
\href{http://arxiv.org/abs/1502.04267}{{\ttfamily arXiv:1502.04267 [hep-th]}}.

\bibitem{Casini2013}
H.~Casini, M.~Huerta, and J.~A. Rosabal, ``{Remarks on entanglement entropy for
gauge fields},'' \href{http://dx.doi.org/10.1103/PhysRevD.89.085012}{{\em
Phys.Rev.} {\bfseries D89} (2014) 085012},
\href{http://arxiv.org/abs/1312.1183}{{\ttfamily arXiv:1312.1183 [hep-th]}}.

\bibitem{Buividovich2008b}
P.~V. Buividovich and M.~I. Polikarpov, ``{Entanglement entropy in gauge
theories and the holographic principle for electric strings},''
\href{http://dx.doi.org/10.1016/j.physletb.2008.10.032}{{\em Phys. Lett.}
{\bfseries B670} (2008) 141--145},
\href{http://arxiv.org/abs/0806.3376}{{\ttfamily arXiv:0806.3376 [hep-th]}}.

\bibitem{Donnelly2011}
W.~Donnelly, ``{Decomposition of entanglement entropy in lattice gauge
theory},'' \href{http://dx.doi.org/10.1103/PhysRevD.85.085004}{{\em
Phys.Rev.} {\bfseries D85} (2012) 085004},
\href{http://arxiv.org/abs/1109.0036}{{\ttfamily arXiv:1109.0036 [hep-th]}}.

\bibitem{Radicevic2014}
\DJ.~Radi\v cevi\'c, ``{Notes on Entanglement in Abelian Gauge Theories},''
\href{http://arxiv.org/abs/1404.1391}{{\ttfamily arXiv:1404.1391 [hep-th]}}.

\bibitem{Soni2015}
R.~M~Soni and S.~P.~Trivedi, ``Aspects of Entanglement Entropy for Gauge Theories,''
\href{http://dx.doi.org/10.1007/JHEP01(2016)136}{\emph{JHEP} \textbf{1601} (2016) 136},
\href{http://arxiv.org/abs/1510.07455}{{\ttfamily arXiv:1510.07455 [hep-th]}}.

\bibitem{vanAcoleyen2015}
K.~van Acoleyen, N.~Bultinck, J.~Haegeman, M.~Marien, V.~B.~Scholz, F.~Verstraete, ``Entanglement of Distillation in Gauge Theories,''
\href{http://arxiv.org/abs/1511.04369}{{\ttfamily arXiv:1511.04369 [quant-ph]}}.

\bibitem{Solodukhin2008}
S.~N.~Solodukhin, ``Entanglement entropy, conformal invariance and extrinsic geometry,''
\href{http://dx.doi.org/10.1016/j.physletb.2008.05.071}{Phys. Lett. B {\bfseries 65} 4 (2008) 305--309},
\href{http://arxiv.org/abs/0802.3117}{{\ttfamily arXiv:0802.3117 [hep-th]}}.

\bibitem{Casini2011}
H.~Casini, M.~Huerta and R.~C.~Myers, ``Towards a Derivation of Holographic Entanglement Entropy,''
\href{http://dx.doi.org/10.1007/JHEP05(2011)036}{\emph{JHEP} \textbf{1105} (2011) 036},
\href{http://arxiv.org/abs/1102.0440}{{\ttfamily arXiv:1102.0440 [hep-th]}}.

\bibitem{Dowker2010}
J.~S.~Dowker, ``Entanglement Entropy for Even Spheres,''
\href{http://arxiv.org/abs/1009.3854}{{\ttfamily arXiv:1009.3854 [hep-th]}}

\bibitem{Casini2015}
H.~Casini and M.~Huerta, ``Entanglement entropy of a Maxwell field on the sphere,''
\href{http://dx.doi.org/10.1103/PhysRevD.93.105031}{ {\em Phys. Rev.} {\bfseries  D93} (2016) 105031},
\href{http://arxiv.org/abs/1512.06182}{{\ttfamily arXiv:1512.06182 [hep-th]}}.

\bibitem{Casini2014}
H.~Casini, M.~Huerta, ``{Entanglement Entropy for a Maxwell Field: Numerical Calculations on a Two-Dimensional Lattice},'' \href{http://dx.doi.org/10.1103/PhysRevD.90.105013}{\emph{Phys.Rev.} \textbf{D90} (2014) 105013}, \href{http://arxiv.org/abs/1406.2991}{{\ttfamily arXiv:1406.2991 [hep-th]}}.

\bibitem{Donnelly2014}
W.~Donnelly, ``{Entanglement Entropy and non-Abelian Gauge Symmetry},''
\href{http://dx.doi.org/10.1088/0264-9381/31/21/214003}{\emph{Class. Quantum Grav.} \textbf{31} (2014) 214003},
\href{http://arxiv.org/abs/1406.7304}{ {\ttfamily arXiv:1406.7304 [hep-th]}}.

\bibitem{Donnelly2014b}
W.~Donnelly, and A.~C. Wall, ``Entanglement Entropy of Electromagnetic Edge Modes,''
\href{http://dx.doi.org/10.1103/PhysRevLett.114.111603}{\emph{Phys. Rev. Lett.} \textbf{114} (2015) 111603},
\href{http://arxiv.org/abs/1412.1895}{ {\ttfamily arXiv:1412.1895} [hep-th] }.

\bibitem{Donnelly2015}
W.~Donnelly, and A.~C.~Wall, ``Geometric Entropy and Edge Modes of the Electromagnetic Field,''
\href{http://arxiv.org/abs/1506.05792}{{\ttfamily arXiv:1506.05792 [hep-th]}}.

\bibitem{Huang2014}
K-W.~Huang, ``Central Charge and Entangled Gauge Fields,''
\href{http://dx.doi.org/10.1103/PhysRevD.92.025010}{{\em Phys. Rev.} {\bfseries D92} (2015) 025010},
\href{http://arxiv.org/abs/1412.2730}{{\ttfamily arXiv:1412.2730 [hep-th]}}.

\bibitem{Zuo2016}
F.~Zuo, ``A note on electromagnetic edge modes,''
\href{http://arxiv.org/abs/1601.06910}{{\ttfamily arXiv:1601.06910 [hep-th]}}.

\bibitem{Radicevic2015}
\DJ.~Radi\v cevi\'c, ``Entanglement in Weakly Coupled Lattice Gauge Theories,''
\href{http://dx.doi.org/10.1007/JHEP04(2016)163}{\emph{JHEP} \textbf{1604} (2016) 163},
\href{http://arxiv.org/abs/1509.08478}{{\ttfamily arXiv:1509.08478 [hep-th]}}.

\bibitem{Kabat1995}
D.~N.~Kabat, ``Black Hole Entropy and Entropy of Entanglement,''
\href{http://dx.doi.org/10.1016/0550-3213(95)00443-V}{Nucl. Phys. B \textbf{453} (1995) 281--302},
\href{http://arxiv.rg/abs/hep-th/9503016}{{\ttfamily arXiv:hep-th/9503016}}.

\bibitem{Solodukhin2011}
S.~N.~Solodukhin, ``Entanglement Entropy of Black Holes,''
\href{http://dx.doi.org/10.12942/lrr-2011-8}{iving Rev. Relativity \textbf{14} (2011) 8},
\href{http://arxiv.org/abs/1104.3712}{{\ttfamily  arXiv:1104.3712 [hep-th]}}.

\bibitem{Cardy2007}
J.~L.~Cardy, O.~A.~Castro-Alvaredo and B.~Doyon, ``Form factors of branch-point twist fields in quantum integrable models and entanglement entropy,''
\href{http://dx.doi.org/10.1007/s10955-007-9422-x}{J. Stat. Phys. \textbf{130} (2008) 129--168},
\href{http://arxiv.org/abs/0706.3384}{{\ttfamily arXiv:0706.3384 [hep-th]}}.

\bibitem{Fursaev2013}
D.~V.~Fursaev, A.~Petrushev and S.~N.~Soldukhin, ``Distributional Geometry of Squashed Cones,''
\href{http://dx.doi.org/10.1103/PhysRevD.88.044054}{{\em Phys. Rev.} {\bfseries D88} (2013) 044054},
\href{http://arxiv.org/abs/1306.4000}{{\ttfamily arXiv:1306.4000 [hep-th]}}.

\bibitem{Kitaev2005}
A.~Kitaev and J.~Preskill, ``Topological Entanglement Entropy,''
\href{http://dx.doi.org/10.1103/PhysRevLett.96.110404}{\emph{Phys. Rev. Lett.} \textbf{96} (2006) 110404},
\href{http://arxiv.org/abs/hep-th/0510092}{{\ttfamily arXiv:hep-th/0510092}}.

\bibitem{Levin2005}
M.~Levin and X-G.~Wen, ``Detecting Topological Order in a Ground State Wavefunction,''
\href{http://dx.doi.org/10.1103/PhysRevLett.96.110405}{\emph{Phys. Rev. Lett.} \textbf{96} (2006) 110405},
\href{http://arxiv.org/abs/cond-mat/0510613}{{\ttfamily arXiv:cond-mat/0510613}}.

\bibitem{Pretko2015}
M.~Pretko and T.~Senthil, ``Entanglement Entropy of $U(1)$ Quantum Spin Liquids,''
\href{http://arxiv.org/abs/1510.03863}{{\ttfamily arXiv:1510.03863 [hep-th]}}.

\bibitem{Grover2011}
T.~Grover, A.~M.~Turner and A.~Vishwanath, ``Entanglement Entropy of Gapped Phases and Topological Order in Three dimensions,''
\href{http:dx.doi.org/10.1103/PhysRevB.84.195120}{ {\em Phys. Rev.} {\bfseries  B84} (2011) 195120}

\bibitem{ArfkenBook}
G.~B.~Arfken, H.~J.~Weber and F.~E.~Harris, \emph{Mathematical methods for physicists: a comprehensive guide}, Academic press, 2011.

\bibitem{Ryu2006}
S.~Ryu and T.~Takayanagi, ``Aspects of Holographic Entanglement Entropy,''
\href{http://dx.doi.org/10.1088/1126-6708/2006/08/045}{\emph{JHEP} \textbf{0608} (2006) 045},
\href{http://arxiv.org/abs/hep-th/0605073}{{\ttfamily arXiv:hep-th/0605073}}.

\bibitem{Lewkowycz2013}
A.~Lewkowycz and J.~Maldacena, ``Generalized Gravitational Entropy,''
\href{http://dx.doi.org/10.1007/JHEP08(2013)090}{\emph{JHEP} \textbf{1308} (2013) 090},
\href{http://arxiv.org/abs/1304.4926}{{\ttfamily arXiv:1304.4926 [hep-th]}}.

\bibitem{GradshteynBook}
I.~S.~Gradshteyn and I.~M.~Ryzhik \emph{Table of integrals, series, and products}, Academic press, 2014.

\bibitem{ParkerBook}
L.~Parker and D.~Toms, \emph{Quantum field theory in curved spacetime: quantized fields and gravity}, Cambridge University Press, 2009.

\end{thebibliography}
\end{document}